\documentclass[twocolumn]{aa} 

\usepackage[varg]{txfonts}

\usepackage[utf8]{inputenc} 
\usepackage{graphicx}	
\usepackage{amsmath}	
\usepackage{amssymb}	
\usepackage{bm}		
\usepackage{epstopdf} 
\usepackage{float}

\usepackage{xcolor}
\usepackage[unicode=true, colorlinks=true, linkcolor=purple, citecolor=blue, filecolor=red, urlcolor=green]{hyperref}

\newcommand{\igr}{\mbox{IGR J18027--2016} }

\newcommand{\hard}{{\em hard} }
\newcommand{\soft}{{\em soft} }

\newcommand{\pwake}{{\em photoionisation wake} }
\newcommand{\awake}{{\em accretion wake} }
\newcommand{\stromgren}{Str{\"o}mgren}

\newcommand{\swift}{{\em Swift}/BAT }
\newcommand{\xmm}{XMM--{\em Newton} }
\newcommand{\integral}{ISGRI/INTEGRAL }

\graphicspath{{./}{figures/}}

\begin{document}

   \title{Stellar wind structures in the eclipsing binary system \igr}

   \author{Federico~A.~Fogantini\inst{1,2}\thanks{fafogantini@iar.unlp.edu.ar}
          \and
          Federico~García\inst{3,4}
          \and
          Jorge~A.~Combi\inst{1,2}
          \and
          Sylvain~Chaty\inst{4,5}
          }

   \institute{Instituto Argentino de Radioastronom\'ia (CCT-La Plata, CONICET; CICPBA; UNLP), C.C. No. 5, 1894 Villa Elisa, Argentina
    \and
    Facultad de Ciencias Astron\'omicas y Geof\'isicas, Universidad Nacional de La Plata, Paseo del Bosque s/n, 1900 La Plata, Argentina
    \and
    Kapteyn Astronomical Institute, University of Groningen, PO BOX 800, NL-9700 AV Groningen, the Netherlands \and
    AIM, CEA, CNRS, Universit\'e Paris-Saclay, Universit\'e de Paris, F-91191 Gif-sur-Yvette, France \and
    Universit\'e de Paris, CNRS, Astroparticule et Cosmologie, F-75013 Paris, France
    }



 
	\abstract
	{IGR J18027--2016 is an obscured high-mass X-ray binary formed by a neutron star accreting from the wind of a supergiant companion with a $\sim$4.57~day orbital period. The source shows an asymmetric eclipse profile that remained stable across several years.}
	{We aim at investigating the geometrical and physical properties of stellar wind structures formed by the interaction between the compact object and the supergiant star.}
	{In this work we analyse the temporal and spectral evolution of this source along its orbit using six archival XMM-{\em Newton} observations and the accumulated {\em Swift}/BAT hard X-ray light curve.}
	{XMM-{\em Newton} light curves show that the source hardens during the ingress and egress of the eclipse, in accordance with the asymmetric profile seen in {\em Swift}/BAT data. A reduced pulse modulation is observed on the ingress to the eclipse. 
	We model XMM-{\em Newton} spectra by means of a thermally-comptonized continuum ({\sc nthcomp}) adding two gaussian emission lines corresponding to Fe~K$\alpha$ and Fe~K$\beta$. We included two absorption components to account for the interstellar and intrinsic media. We found that the local absorption column outside the eclipse fluctuates uniformly around $\sim$ 6$\times$10$^{22}$~cm$^{-2}$, whereas, when the source enters and leaves the eclipse, the column increases by a factor of $\gtrsim$3, reaching values up to $\sim$35 and $\sim$15$\times 10^{22}$~cm$^{-2}$, respectively. }
	{Combining the physical properties derived from the spectral analysis, we propose a scenario where a {\em photo-ionisation wake} (mainly) and an {\em accretion wake} (secondarily) are responsible for the orbital evolution of the absorption column, the continuum emission and the variability seen at the Fe-line complex.}
	{}

	\keywords{X-ray: individual objects: IGR J18027--2016 -- high mass stars -- neutron stars -- X-ray: binaries}

   \maketitle
%
\section{Introduction} \label{sec:intro}

Since its launch, back in 2002, the IBIS/ISGRI detector \citep{Lebrun2003, Ubertini2003} on board the INTEGRAL observatory \citep{Winkler2003} has discovered a large number of hard and obscured X-ray sources. Most of them have a Galactic origin and belong to the class of High Mass X-ray Binaries (HMXB). Based on the spectral type of the companion star (mainly seen in optical), the accretion mechanisms taking place, and their X-ray behaviour, HMXBs are further classified into Be (later called BeXBs) or Supergiant X-ray binaries (SgXBs). In Be systems, the compact object is mainly a neutron star (NS) with moderately eccentric orbits ($P \sim 10$ to $100$~days) and spends short time intervals in close proximity to the dense circumstellar disk surrounding the Be companion \citep{Negueruela2002, Liu2000}. In SgXBs, the compact objects are typically in shorter (1--10~days) orbits around an OB supergiant companion. In those systems, accretion can be driven by Roche-lobe overflow and/or through the powerful supergiant stellar wind \citep[for a review see e.g.][]{Chaty2013}.

In this last category, two sub-classes of binary systems appear. The first one usually exhibits high levels of obscuration ($N_{\rm H} \sim 10^{23} - 10^{24}$~cm$^{-2}$), local to the source \citep{Walter2006, Chaty2008}. 
The second one is composed by binary systems consisting of compact objects associated with a supergiant donor that undergo fast transient outbursts in the X-ray band. These latter sources are called supergiant fast X-ray transients (SFXT) \citep{Negueruela2006, Sguera2006}.

The obscured source \mbox{IGR J18027--2016} is a typical SgXB, which was discovered during the first year of operations by INTEGRAL. The source was spatially associated with the X-ray pulsar \mbox{SAX J1802.7--2017} by \cite{Augello2003}. Later on, \cite{Hill2005} reported that the source is an eclipsing HMXB system composed of an accreting X-ray pulsar and a late OB supergiant star, with persistent emission and high intrinsic photoelectric absorption. Timing analysis performed on the ISGRI data by these authors confirmed the nature of the source, and constrained its orbital period to $P_{\rm orb}=4.5696\pm0.0009$~days with a mid-eclipse ephemeris at $T_{\rm mid}=52931.37\pm0.04$~MJD. 

Its NIR counterpart (\mbox{2MASS J18024194--2017172}) was identified by \cite{Masetti2008}. They obtained a low-resolution optical spectrum and showed that it is a very reddened early-type star. Using the mass and radius deduced by \cite{Hill2005}, they argue that the counterpart should be a B-type giant. Later, \cite{Chaty2008} confirmed the identification of the counterpart. Based on the $R_{\star}/D_{\star}$ value of the normalization that minimizes $\chi^2$ on the blackbody SED fitting to the optical and NIR/MIR data, and assuming a typical B-star radius of $R_{\star} = 20$~$R_{\odot}$,  \cite{Torrejon2010} estimated a distance to the source of $D_{\star}=$12.4$\pm$0.1~kpc within a 90\% confidence level.

Using \swift and \integral  data, \cite{Coley2015} and \cite{Falanga2015} independently provided better constraints to \igr parameters such as orbital period and eclipsing phases. They also proposed different mechanisms that may be responsible for creating an asymmetric eclipse profile, including large stellar wind structures (such as accretion wakes and photo-ionization wakes) and non-zero eccentricity orbits (0.04$<$e$<$0.2). 
With a different perspective, \cite{Aftab2019} and \cite{Pradhan2019} studied the orbital behaviour of \igr using \swift and \xmm data, and they concluded that stellar wind clumps may be responsible for the short and long term variability and spectral behaviour of the source.

In this paper, we present a detailed temporal and spectral analysis of all publicly available \xmm and \swift observations (same data set as \cite{Pradhan2019}) to investigate the behaviour of the X-ray emission of \igr along the orbital revolution of the system. In our approach, we specially consider and treat the strong background and pile--up that affects half of the \xmm observations.
In \hyperref[sec:reduction]{Section~\ref{sec:reduction}}, we provide information on the observations and data reduction methods employed for the analysis. 
We describe the results of our temporal and spectral X-ray analysis in \hyperref[sec:results]{Section~\ref{sec:results}}. 
In \hyperref[sec:discussion]{Section~\ref{sec:discussion}}  we discuss our results and introduce a possible astrophysical scenario to describe the temporal and spectral orbital variability of the system. 
Finally, a summary of all the obtained results is presented on \hyperref[sec:conclusions]{Section~\ref{sec:conclusions}}.

%
\section{Observations and data reduction} \label{sec:reduction}

\subsection{XMM--{\em Newton} data}

The XMM-{\em Newton} observatory contains two X-ray instruments: the European Photon Imaging Camera (EPIC) and the Reflecting Grating Spectrometers (RGS). EPIC is formed by three detectors, a PN camera \citep{Struder2001} and two MOS cameras \citep{Turner2001}, all operating in the 0.3$-$12~keV energy range. RGS consists of two high-resolution spectrometers sensitive  in the 0.3$-$2.0~keV soft energy range.

XMM-{\em Newton} first pointed at \igr on April 06, 2004 (PI: R. Walter) and then five times more between September 06 and September 13, 2014 (PI: A. Manousakis). The first observation was performed with a Medium filter in Large Window mode, while the rest of the exposures were conducted with a Medium filter in Full Frame observation mode. As RGS covers only the highly absorbed soft band up to $\sim$2.0~keV, we are prevented of using those spectrometers in our analysis, given the highly obscured nature of \igr. 
For our analysis, we started considering PN and MOS cameras, however, due to the larger integration time of MOS cameras and the presence of pile--up (see below), in what follows we only consider the PN camera.

We reduced the XMM--{\em Newton} data using the Science Analysis System (SAS) version 18.0.0 and the latest calibrations available by February 2019. By processing the Observation Data Files (ODF) with the {\sc epproc} task we produced event lists from the PN data sets. Basic information on the six observations used in this work is given in \hyperref[tab:obs]{Table~\ref{tab:obs}}.

\begin{table*} 
\centering
\begin{tabular}{c c c c c c c c c c c c}
\hline
\hline
Id & OBSID & Start date & Filter/Mode   & Exposure    & Phase & GTI (r$<$0.4)   & GTI (r$<$1)  & Exc. Radius \\
   &        & (UTC)     &               & (ks)              &       & (ks)        & (ks)     & (PhU) \\
\hline
26 & 0206380601 & 2004-04-06 \, 06:34 & Medium/LW & 10.2 &   0.188$\pm$0.009  &  7.3  &  --  &  0  \\
\hline
77 & 0745060701 & 2014-09-06 \, 09:55 & Medium/FW & 14.3 &  -0.155$\pm$0.016  & 12.7  & --   &  0 \\
75 & 0745060501 & 2014-09-08 \, 19:58 & Medium/FW & 16.0 &   0.377$\pm$0.018  & 14.3  & --   & 100 \\
76 & 0745060601 & 2014-09-09 \, 22:59 & Medium/FW & 17.0 &  -0.376$\pm$0.019  & 15.2  & --   & 120 \\
74 & 0745060401 & 2014-09-11 \, 22:30 & Medium/FW & 43.0 &   0.109$\pm$0.053  &  3.3  & 33.3 &  0 \\
78 & 0745060801 & 2014-09-12 \, 19:28 & Medium/FW & 16.5 &   0.237$\pm$0.008  &  6.9  & --   & 100 \\
\hline
\end{tabular}
\caption{XMM-{\em Newton} PN observations used in this work. LW and FW correspond to Large Window and Full Window modes respectively. Phases were calculated using a period of 4.56993~days and a mid-eclipse reference time of 55083.82~BMJD. Phase errors correspond to observations exposure time after GTI filtering. {\em Id} column denotes an abbreviated form of the OBSID which will be used to refer to them. GTI columns indicate the remaining ks of the total PN exposure times for background 10-12~keV light curve upper rate limit of 0.4~cps and 1~cps, respectively. The last column contains the excision radii in physical units (PhU) adopted after the pile-up analysis. }
\label{tab:obs}
\end{table*}


\subsubsection{Good-time intervals and pile-up treatment} \label{sec:gti}

In order to exclude high-background periods and generate good-time intervals (GTI) we produced light curves of 100~sec bin excluding the entire CCD ({\sc CCDNR!=4}) in which the bright \igr source is located. For them we only considered events with energies in the 10--12~keV range. 

\begin{figure}
\centering
\includegraphics[width=\columnwidth, angle=0, clip]{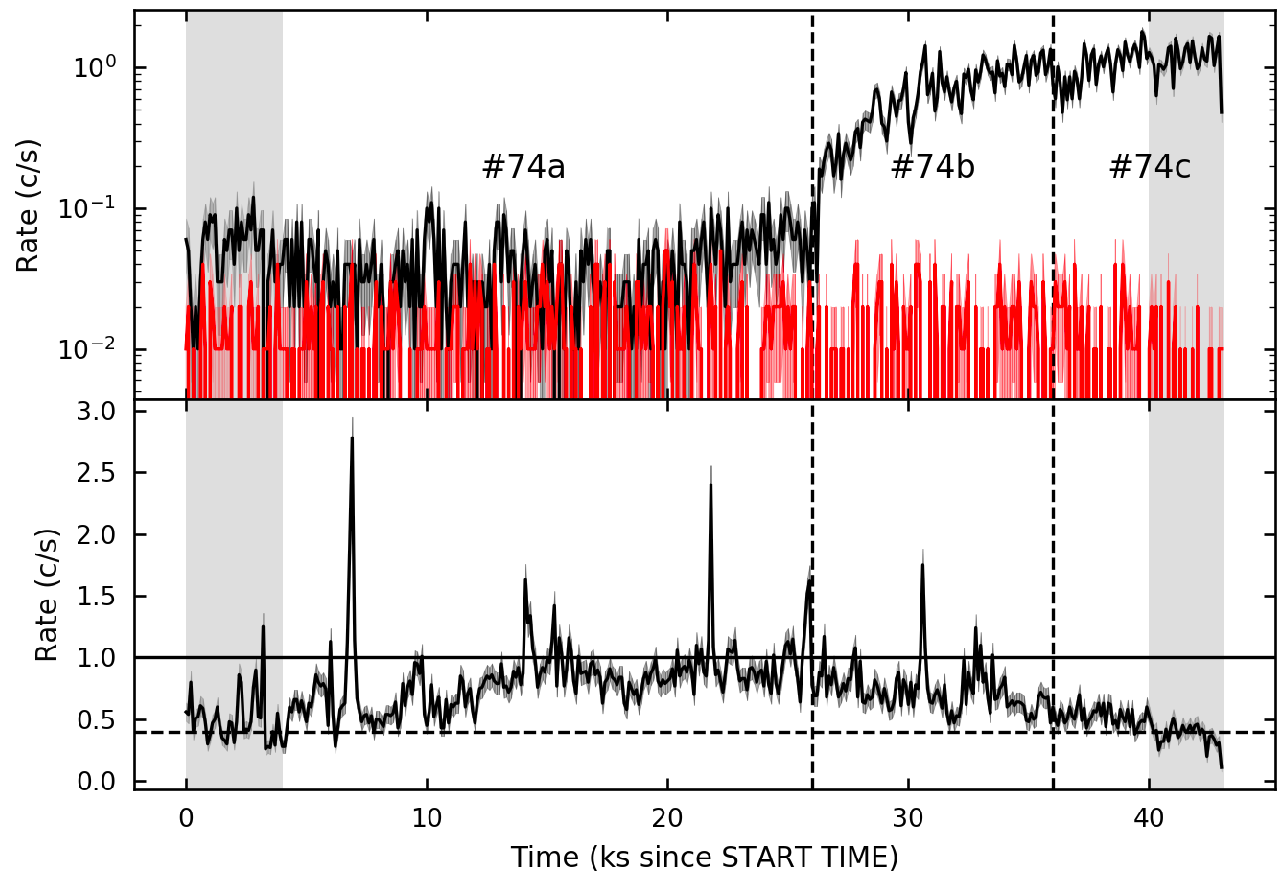}
\caption{
Source and background light curves of the eclipse observation \#74 using 100~sec bins. Upper panel: source+background (black) and background (red) light curves extracted in the 0.5--12 keV energy range using circular regions with equal area. Bottom panel: full CCD light curve in the 10--12 keV energy range excluding the source region. Black vertical lines indicate split times for eclipse (a), egress (b) and out-of-eclipse (c) periods according to \citep{Coley2015} ephemeris. The full (dashed) horizontal line corresponds to the background rate of 1~cps (0.4~cps).
}
\label{fig:gti74}
\end{figure}

Some of the PN exposures were heavily affected by high background activity. In such cases, following the standard GTI procedure (rate $<$ 0.4 cps), the effective exposure times were strongly reduced, which avoided further scientific analysis. In particular, for the observation covering the eclipse (\#74), from the total $\sim$40~ks of exposure time, only 7\% is retained after filtering. This particular example is shown on \hyperref[fig:gti74]{Figure~\ref{fig:gti74}}. The entire collection of background lightcurves can be found on \hyperref[fig:pnbkglcs]{Figure~\ref{fig:pnbkglcs}} on the \hyperref[sec:appendix]{Appendix~\ref{sec:appendix}} on the online version of this work.
 
The upper panel of \hyperref[fig:gti74]{Figure~\ref{fig:gti74}} shows the source-plus-background (black) and background only (red) light curves at 0.5--12~keV (100~sec bin) for different circular regions of the same radius (600 physical units, or PhU). Lower panel shows the full-CCD background light curve in the 10--12~keV (100~sec bin) energy range, excluding the source region. When following the standard cut-out rate of 0.4~cps (horizontal dashed line on the bottom panel), only a small portion of the first and last 3~ks (vertical grey stripes) remains in the GTI. Thus, in such a case, the full information associated to the eclipse and its egress would be lost. Taking into account the high count rate exhibited by the source in all the observations, we decided to increase the standard rate limit for GTIs to 1~cps, in order to minimize the issue raised by the presence of the background. With this new limit, bright and fast background flares are still excluded, while the exposure times for science are highly increased, and the background count-rate in the 0.5--12~keV energy range for a region with the same area as the source extraction region is negligible, with respect to the source events (see red lightcurve on \hyperref[fig:gti74]{Figure~\ref{fig:gti74}}) . 

In \hyperref[tab:obs]{Table~\ref{tab:obs}} we present the remaining exposure times (in kiloseconds) of each observation, according to the chosen background rate limit. One column shows the GTIs for the standard rate of 0.4~cps while the second column corresponds to the selected limit of 1~cps. To be conservative we used the increased rate limit only for eclipse observation \#74.

In order to check for {\em pile-up} effects, we used the SAS task {\sc epatplot} to create diagnoses of the relative ratios of single- ({\sc PATTERN==0}) and double- ({\sc PATTERN in [1:4]}) events and to study their deviations from the expected calibrated values. Using a circular region of 600 PhU centred at the source and 0.5--12~keV energy range ({\sc pileupnumberenergyrange}), we found that 3 of the 6 observations were affected by pile-up. 

For each of the observations showing pile-up we produced new {\sc epatplots} extracting events from a series of annuli varying the inner radius of the PSF by 10, 20, 30, etc PhU, with a fixed outer radius of 600 PhU. We then searched for the minimum excision inner radius in which the pattern coefficients for {\em single} and {\em double} (reported on the {\sc epatplots}) were consistent with unity within errors. In the last column of \hyperref[tab:obs]{Table~\ref{tab:obs}}, we show the excision radii considered for each of the observations.

We note here that, in contrast, \cite{Pradhan2019} mention that the high energy background above the standard count rate limit only affected observation \#26 and that none of the observations is affected by pile-up. 
A strict comparison is odd due to the absence of detailed information on the background treatment.
Both light-curves and spectra might be affected by both background and pile-up effects.



\subsection{{\em Swift}/BAT data}

The {\em Swift}/Burst Alert Telescope (BAT) is an X-ray instrument working in the 15--50~keV energy range which provides almost real-time monitoring of the hard X-ray sky. BAT covers $\sim$90\% of the sky each day reaching full-day detection sensitivities of 5.3 mCrab, at a temporal resolution of 64~s \citep{Krimm2013}. 

In this paper we consider the full daily light curve of \igr available up to February 14, 2019 in the {\em Swift}/BAT service\footnote{\href{https://swift.gsfc.nasa.gov/results/transients/weak/IGRJ18027-2016/}{https://swift.gsfc.nasa.gov/results/transients/weak/IGRJ18027-2016/}}, a public website where more than 900 light curves of hard X-ray sources are available, spanning for more than 8 years.


\section{Results} \label{sec:results}

\subsection{X-ray light curves} \label{sec:lightcurves}

\subsubsection{{\em Swift}/BAT}

After retrieving the full light curve, we created the orbital-period folded light curve using 50 phase bins based on the refined ephemeris of \citet{Coley2015} ($P_{\rm orb}$=4.56993~days and $T_{\rm mid}$=55083.82~BMJD). In this paper we will consider mid eclipse phase to be $\phi=0$.

\begin{figure}
\centering
\includegraphics[width=\columnwidth, clip]{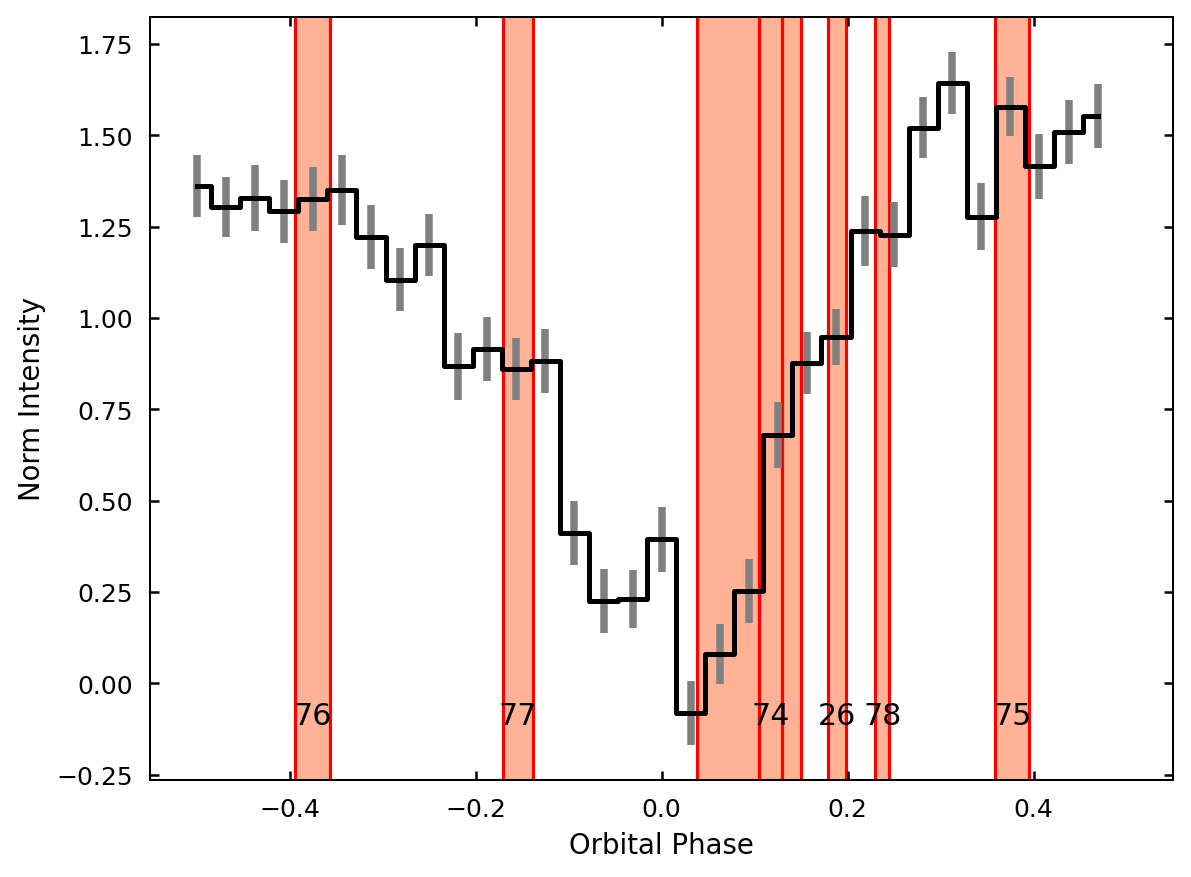}
\includegraphics[width=\columnwidth, clip]{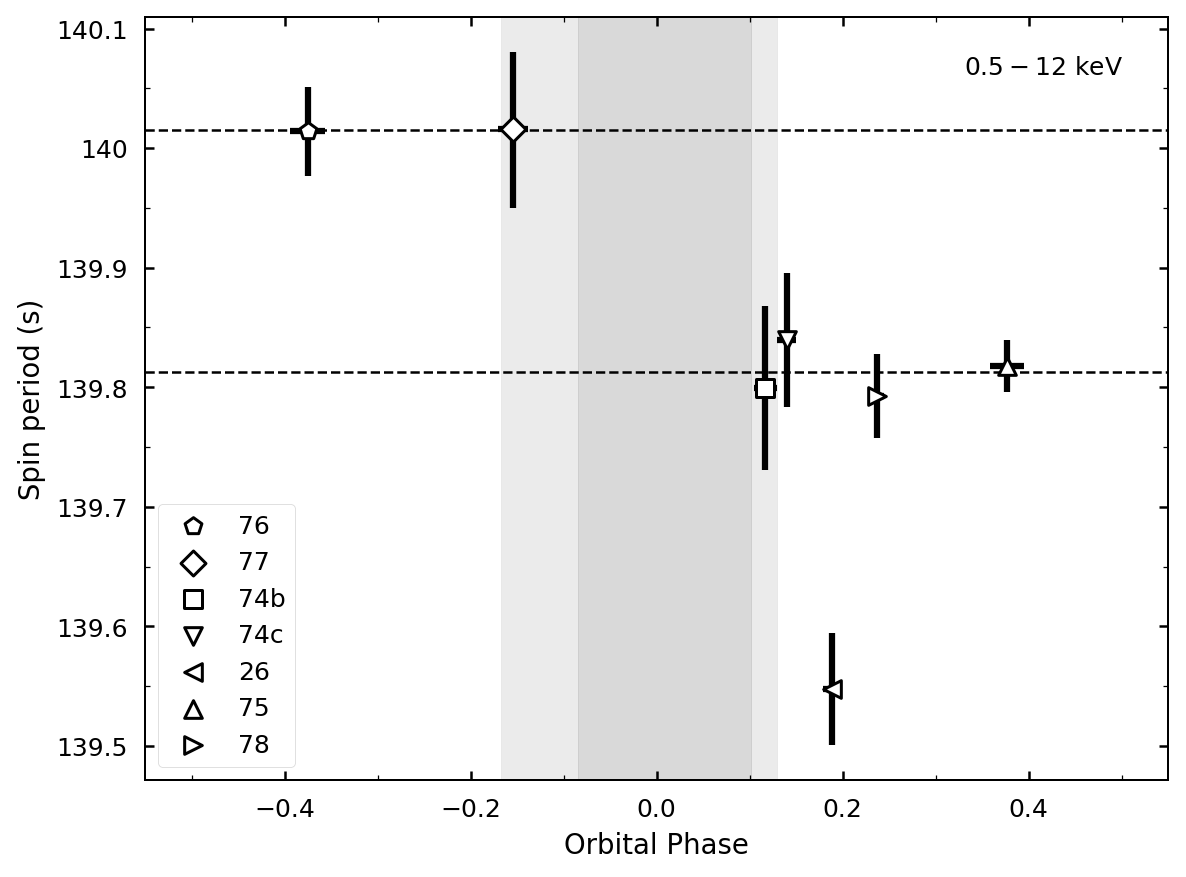}

\caption{Top: Folded {\em Swift}/BAT daily light curve using 32 phase bins with a period of 4.56993~days and mid-eclipse reference time of 55083.82~BMJD \citep{Coley2015}. The vertical light red stripes correspond to the six XMM-{\em Newton} observations used on this paper with their widths representing their final exposure times after GTI filtering.
Bottom: NS spin periods for each observation, estimated by fitting a Lorentzian profile to the $\chi^2$ distribution of the period search. Observation~\#74a (eclipse) was not included on the analysis. We register average spin periods of 140.015$\pm$0.0008~s and 139.81$\pm$0.02~s before and after the eclipse respectively (excluding Obs~\#26, as it corresponds to a different epoch). Averages are indicated with dashed lines.}  
\label{fig:spins}
\end{figure}

From \hyperref[fig:spins]{Figure~\ref{fig:spins}} one can clearly identify the eclipsing-binary nature of the source, where the eclipse spans for $\sim$20\% of the entire cycle. Moreover, it can also be noticed that the transition into the eclipse is more gradual than the eclipse egress. This corresponds to the asymmetry on the eclipse profile already studied by \cite{Coley2015} and \cite{Falanga2015}.

\subsubsection{XMM-{\em Newton}}

We extracted PN light curves using {\sc evselect} task from SAS using 1, 10, 50 and 100~sec bin sizes for a source region of 30~arcsec (600 PhU). For the background region, we used another circular region of 30~arcsec on the same chip as the source, following XMM--{\em Newton} technical recommendations\footnote{\href{http://xmm2.esac.esa.int/docs/documents/CAL-TN-0018.pdf}{http://xmm2.esac.esa.int/docs/documents/CAL-TN-0018.pdf}}. Background substraction was made by means of {\sc epiclccorr} task from SAS.
Lightcurves were extracted on three energy bands: {\em soft}: 0.5--6~keV, {\em hard}: 6--12~keV and {\em full}: 0.5--12 ~keV for timing analysis and on 0.5--6.1~keV, 6.1--6.7~keV, 6.7--12~keV for further pulse-fraction analysis.

\begin{figure*}
\centering
\includegraphics[width=\columnwidth, clip]{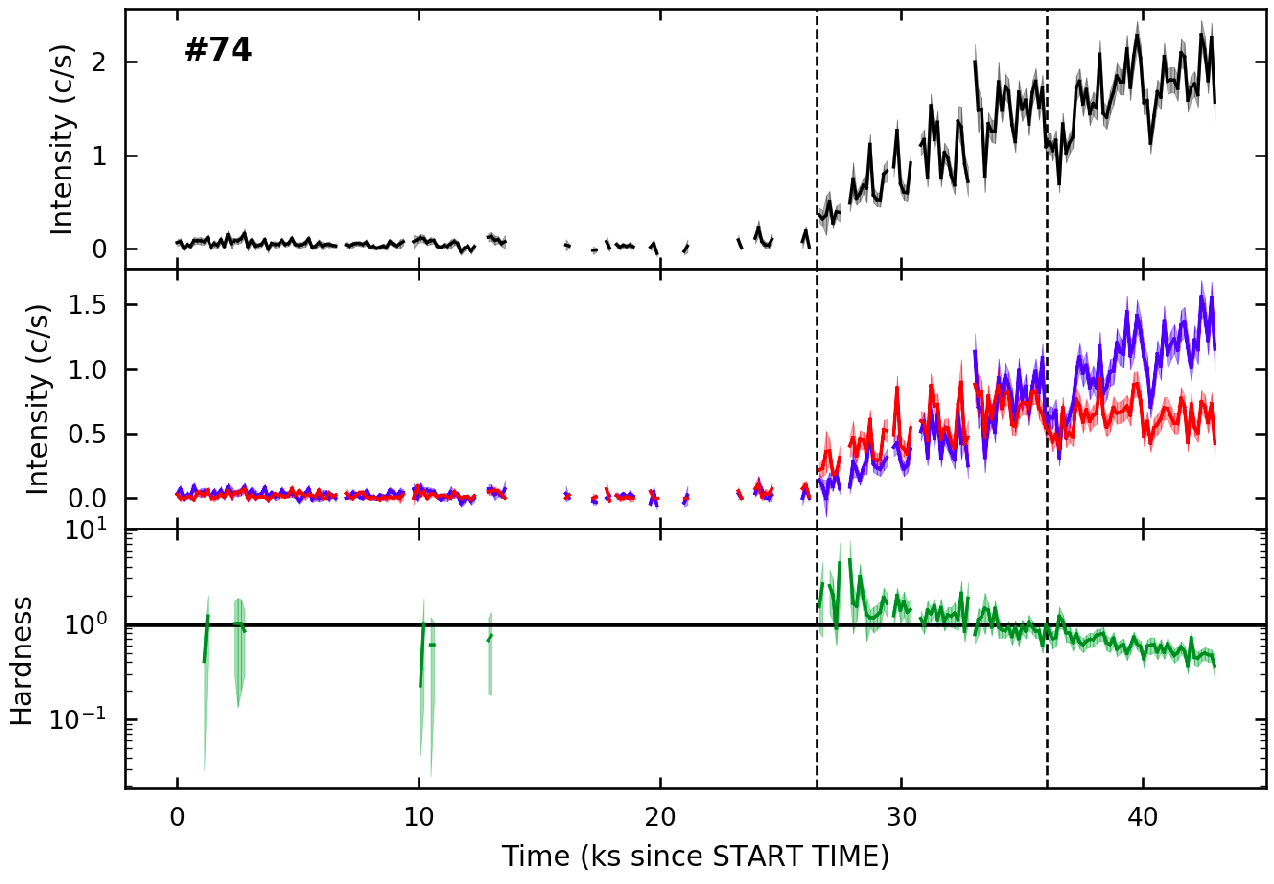}
\includegraphics[width=\columnwidth, clip]{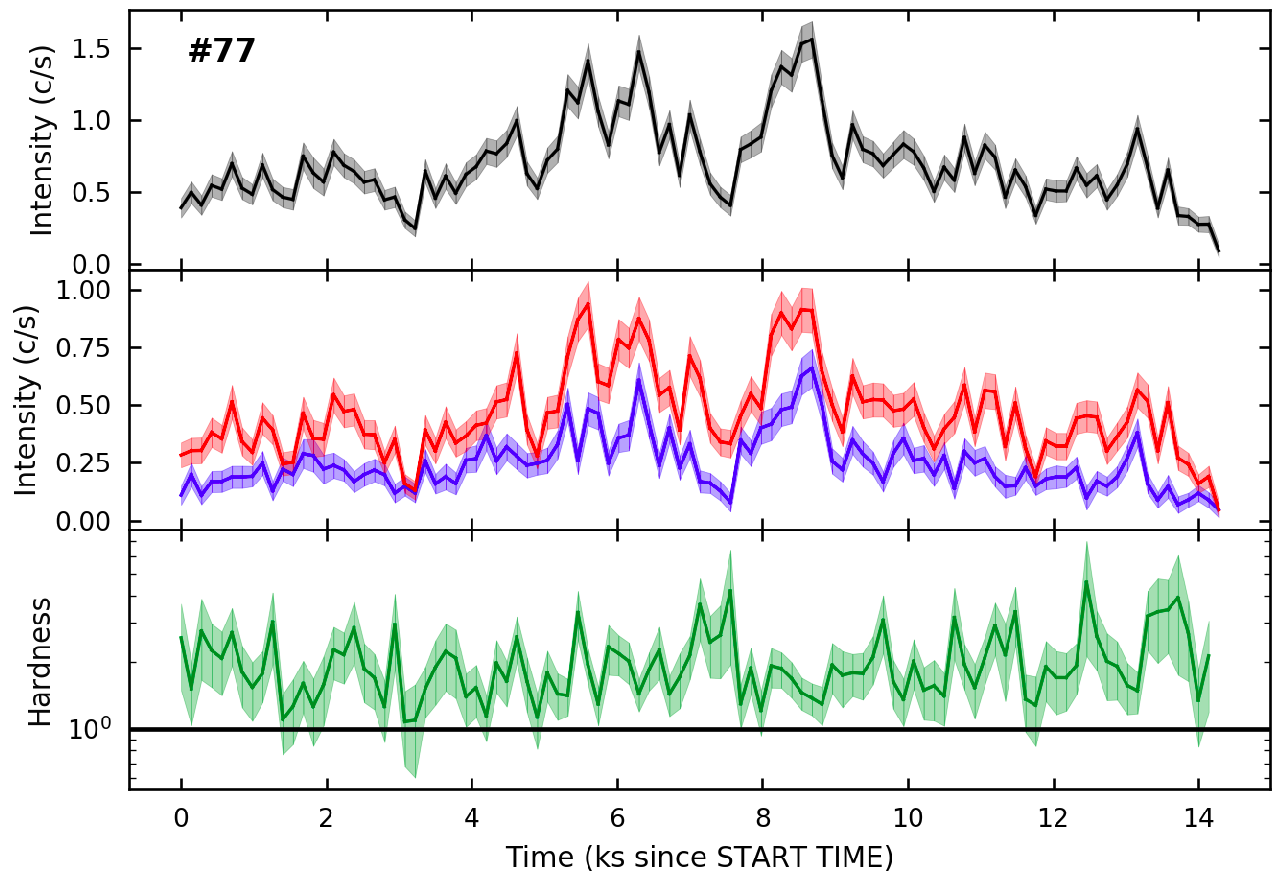}
\caption{{\em XMM--Newton} light curves of observations \#74 (left) and \#77 (right) with 100~sec bin size. Each plot contains 3 panels: {\em full} light curve (upper panel), {\em soft} (blue) and {\em hard} (red) light curves (middle panel) and hardness ratio (green; lower panel). Data gaps correspond to GTI filters and high significance selection (data$>$error).}
\label{fig:xmmlcs}
\end{figure*}

For each observation, we created hardness and intensity averages between {\em soft} and {\em hard} spectral bands. In \hyperref[tab:colors]{Table~\ref{tab:colors}} we present the results found for each individual exposure using 50~s bin light curves. 
In \hyperref[fig:xmmlcs]{Figure~\ref{fig:xmmlcs}} we show light curves with 100~s bins of observations \#74 (egress of eclipse) and \#77 (ingress to eclipse). Each plot contains three panels: top for 0.5--12~keV light curve ({\em full}; black); middle for 0.5--6~keV ({\em soft}; blue) and 6--12~keV ({\em hard}; red) light curves; and bottom for {\em hard}/{\em soft} ratios (green). Data gaps on {\em full}, {\em soft} and {\em hard} light curves of Obs.~\#74 correspond to background-flaring activity shown on bottom panel of \hyperref[fig:gti74]{Figure~\ref{fig:gti74}}. We excluded the hardness-ratio data points that had low significance (data$<$error) which explains the data gaps on this panel.

The top panel in the left plot of \hyperref[fig:xmmlcs]{Figure~\ref{fig:xmmlcs}} shows the eclipse and the egress transition denoted by the two vertical dashed lines ($time$ between $\sim$26~ks and $\sim$36~ks). The time interval of $\sim$10~ks of the eclipse egress we estimated from \hyperref[fig:xmmlcs]{Figure~\ref{fig:xmmlcs}} corresponds to a phase interval of $\sim$0.025, which is in agreement with that of \cite{Coley2015} of 0.027$^{+0.004}_{-0.006}$.

After the eclipse egress (\#74c) the \hard count rate stays almost constant around $\sim$0.5~cps while the \soft rate continues rising, leading to a hardness-ratio (HR) evolution.
We performed a linear fit to the hardness ratio ($\log_{10}HR=At+B$) for the egress transition, obtaining \mbox{$A=-0.060\pm0.003$~cps~ks$^{-1}$} and \mbox{$B=2.99\pm0.14$~cps~ks$^{-1}$} with a reduced $\chi^2=0.84$. 
Using this linear expression we estimated the time where HR$<$1 (within errors), which resulted in $\sim$36~ks, with a corresponding phase in agreement with the egress-phase calculations ($\phi_e + \Delta \phi_e$) of  \cite{Coley2015}.
 
In the right plot of \hyperref[fig:xmmlcs]{Figure~\ref{fig:xmmlcs}},we present observation \#77. During the entire exposure time, the \hard emission dominates over the \soft emission. The hardness ratio stays somewhat constant around $\sim$2. This behaviour is similar to that seen during the ingress of eclipse (\#74b), where the \hard emission dominated over the \soft one. 
In \hyperref[tab:colors]{Table~\ref{tab:colors}} we report the mean values and their standard deviations for every energy band and for all observations except for \#74, as a linear fit was done. As can be seen from these values, for observations outside the eclipse, although \hard emission is higher than that of the eclipse (\#74 and \#77), \soft emission is much higher and dominates at every orbital phase.

In \hyperref[tab:colors]{Table~\ref{tab:colors}} we also included the absorption column density $N_{\rm H}$ in units of 10$^{22}$~cm$^{-2}$. We find a positive correlation between the hardness ratio and the absorption column as expected. Mind that the latter one is derived from spectral fitting and thus model dependent (see \hyperref[sec:spectras]{Section~\ref{sec:spectras}} for details) while lightcurve colours are sensitive only to the instruments response.

\begin{table}
\centering
\begin{tabular}{c c c c c}
\hline
\hline
OBSID  & Soft rate & Hard rate & Hard/Soft ratio & $N_{\rm H}$ \\
\hline
76  &  1.29$\pm$0.50  &  0.97$\pm$0.27  &  0.73$\pm$0.16 & 5.48$^{+0.31}_{-0.23}$ \\
77  &  0.23$\pm$0.11  &  0.43$\pm$0.18  &  1.77$\pm$0.54 & 36.06$^{+3.66}_{-2.02}$ \\
74a &  0.02$\pm$0.03  &  0.38$\pm$0.29  &  1.01$\pm$0.28 & 0.17$^{+0.97}_{-0.14}$ \\
74b &  --  &  --  &  -- & -- \\
74c &  --  &  --  &  -- & -- \\
26  &  1.60$\pm$0.45  &  0.69$\pm$0.19  &  0.43$\pm$0.05 & 8.0$^{+0.7}_{-0.5}$ \\
78  &  2.58$\pm$0.41  &  0.80$\pm$0.14  &  0.30$\pm$0.05 & 0.99$^{+0.08}_{-0.07}$ \\
75  &  1.37$\pm$0.46  &  0.64$\pm$0.14  &  0.45$\pm$0.12 & 2.25$^{+0.09}_{-0.13}$ \\
\hline
\hline
\end{tabular}
\caption{Averaged count rates (in cps) of PN exposures on soft (0.5--6~keV) and hard (6--12~keV) energy bands and hard/soft ratio using 100~s bin light curves. Observations \#74b and \#74c were not averaged but rather linearly modeled (see text). Local absorption column ($N_{\rm H}$, in units of 10$^{22}$~cm$^{-2}$) is included for comparison. Details on $N_{\rm H}$ spectral fitting are described in \hyperref[sec:spectras]{Section~\ref{sec:spectras}}.}
\label{tab:colors}
\end{table}

Using 1~s bin lightcurves in the full energy range (0.5--12~keV) we performed spin-period searches using {\sc efsearch} task from HEASoft. This task returns a $\chi^2$ distribution of fitted periods in a given interval and resolution to the input light curve. 
We then proceeded to fit a Lorentzian profile to the maximum peak found to estimate the best-fit orbital-period and its uncertainty. According to \cite{Larsson1996}, the period error is overestimated by a factor of $\sim$20 with this method. Observation~\#74a (eclipse) was not included on the analysis. These results can be seen in \hyperref[fig:spins]{Figure~\ref{fig:spins}}.

A Doppler shift in the NS spin can be suggested before and after the eclipse (denoted by the dark-grey vertical stripe in \hyperref[fig:spins]{Figure~\ref{fig:spins}}). Before the eclipse we register a weighted average of 140.015$\pm$0.0008~s, higher than the corresponding 139.81$\pm$0.02~s found after the eclipse (excluding observation \#26, which corresponds to a different epoch). 
The NS spin is relatively larger (red-shifted) before the eclipse, when the NS is moving away from the observer, and shorter (blue-shifted) when it goes out of eclipse, moving towards the observer, and thus compatible with the expected Doppler shifts due to the short-period orbital motion of the NS around the high-mass stellar companion (for a high-inclination system like this eclipsing binary).




\subsection{X-ray spectra} \label{sec:spectras}

\begin{figure*}
\centering
\includegraphics[width=\columnwidth, clip]{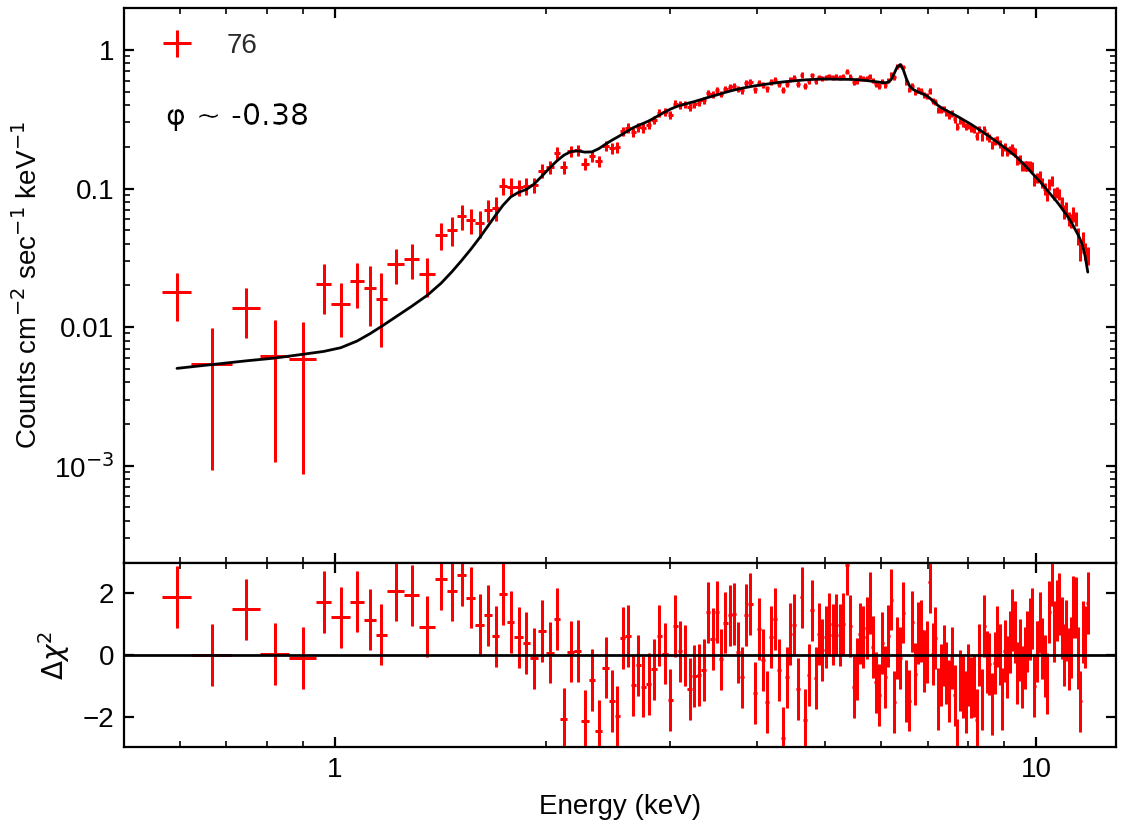}
\includegraphics[width=\columnwidth, clip]{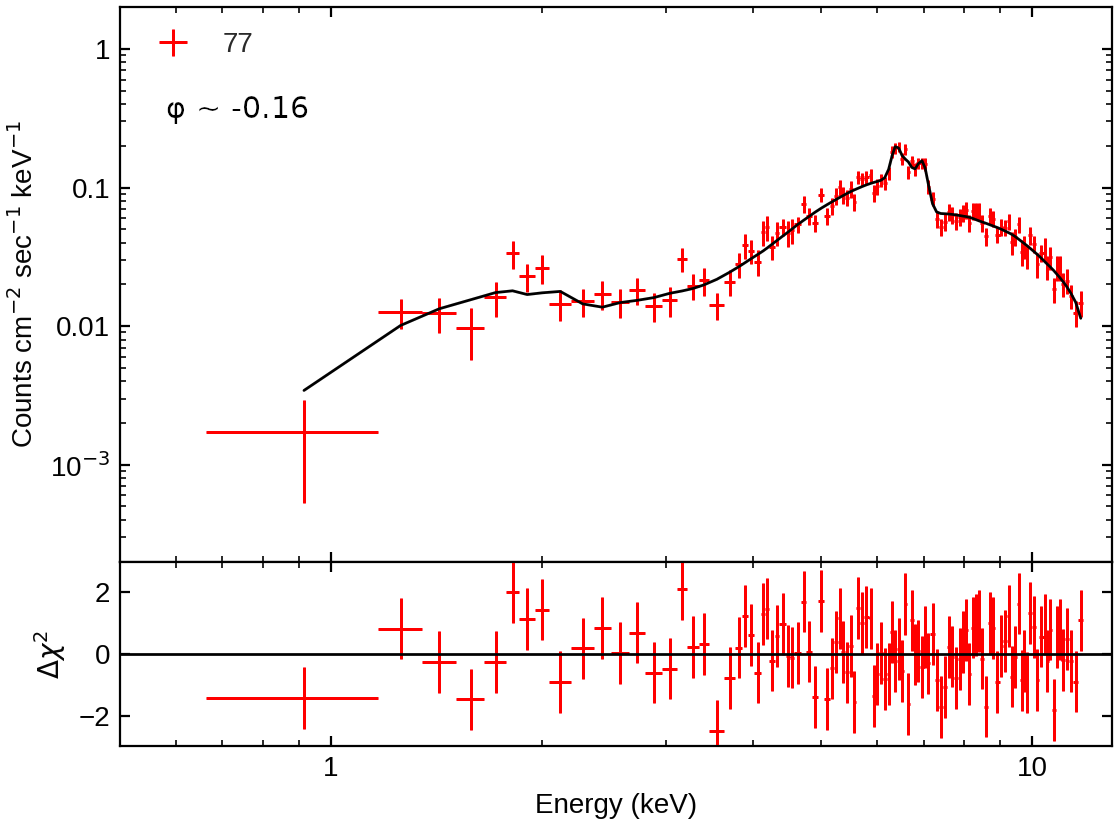}
\includegraphics[width=\columnwidth, clip]{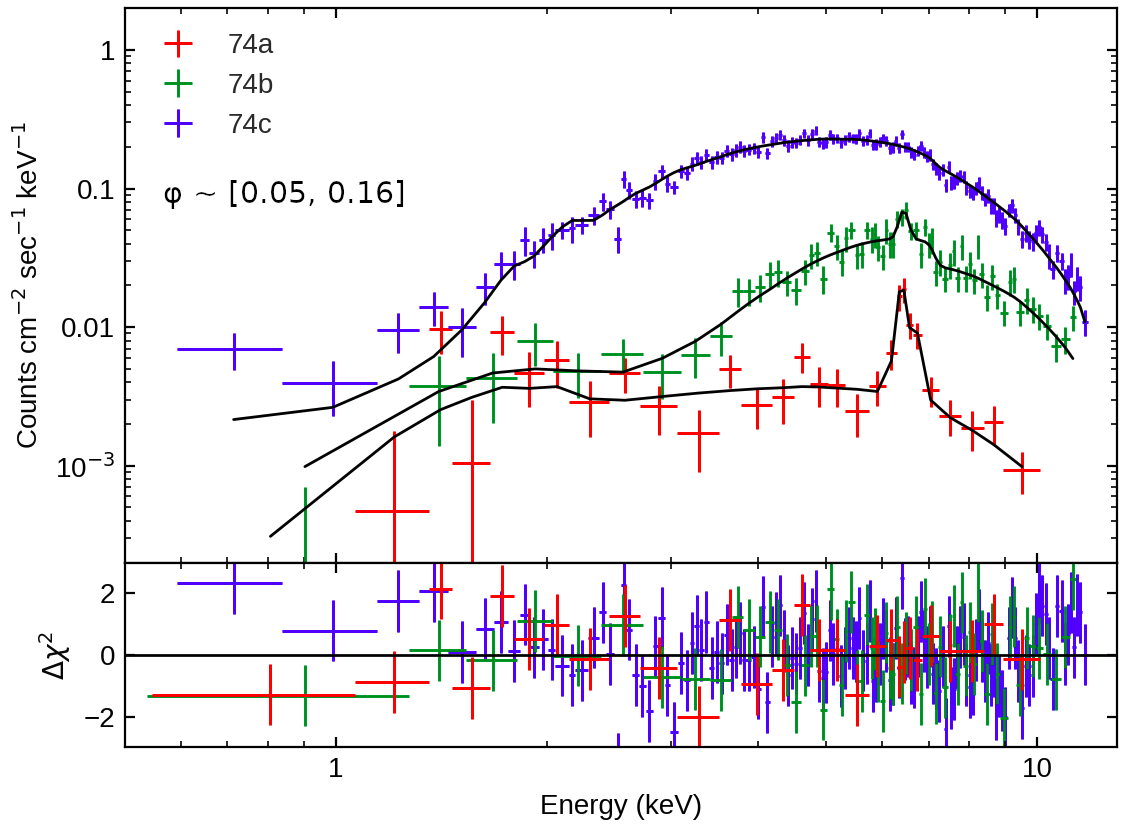}
\includegraphics[width=\columnwidth, clip]{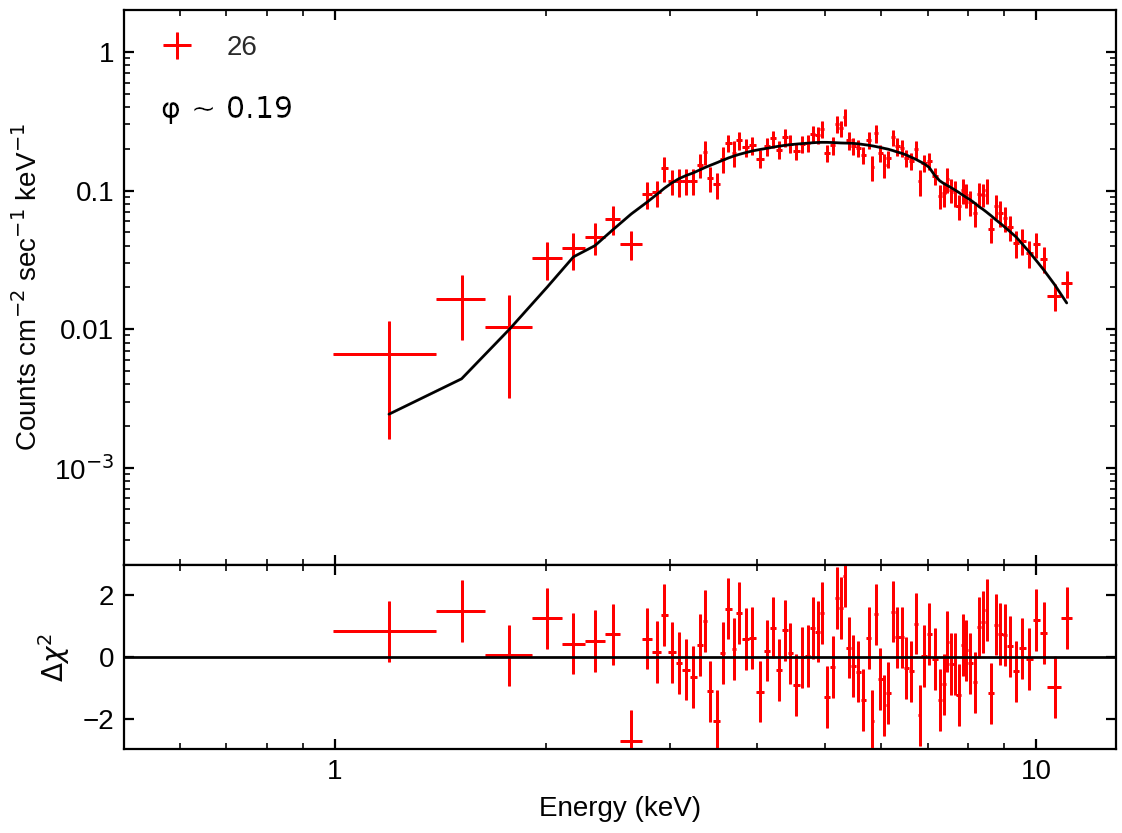}
\includegraphics[width=\columnwidth, clip]{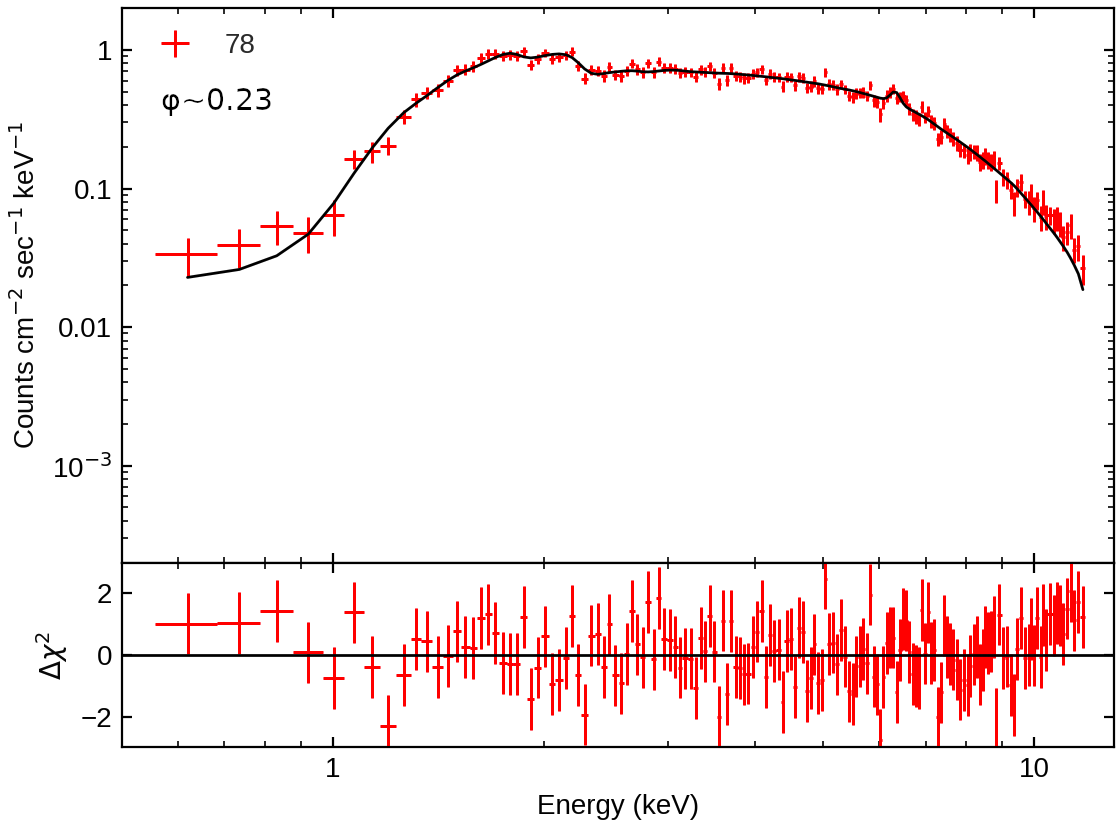}
\includegraphics[width=\columnwidth, clip]{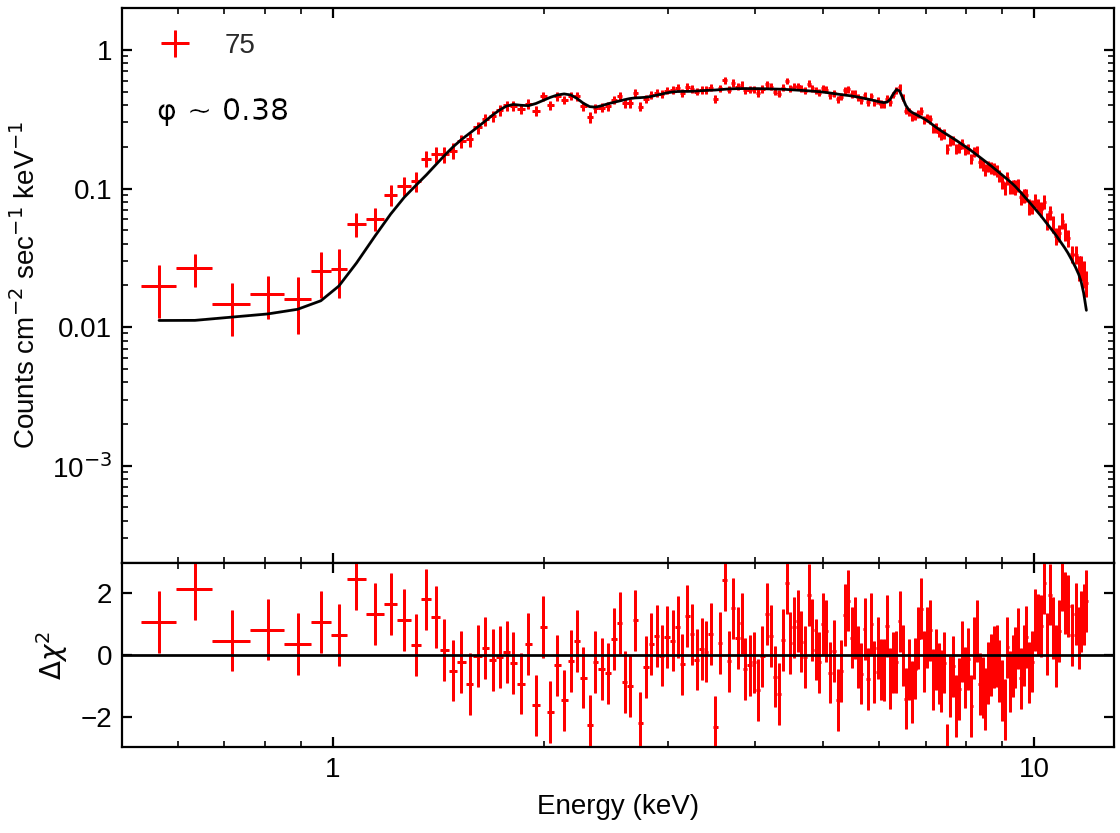}
\caption{The 6 spectra fitted in this work. We used a doubly absorbed {\sc nthcomp} continuum model plus 2 gaussian emission lines according to \hyperref[tab:lines]{Table~\ref{tab:lines}}. Observation \#74 was split onto 3 time periods: \#74a (eclipse), \#74b (egress) and \#74c (out-of-eclipse). Orbital phases ($\phi$) are indicated in each panel.}
\label{fig:specs}
\end{figure*}

We extracted X-ray spectra in the same annular regions determined by the analysis of the {\sc epatplot} (see previous section) with an outer radius of 600 PhU. We used the same background regions indicated in the light-curve analysis. We generated redistribution response matrices (RMF) using {\sc rmfgen}, and ancillary response files (ARF) via {\sc arfgen} tasks, respectively. Spectra were binned to 30 counts per bin with an over-sample factor of 3. Simple and double pattern events were selected.


In general, the spectra of IGR J18027--2016 can be well described by an absorbed power-law like continuum with high energy cut offs at energies above 7~keV. Features such as a soft excess at energies below 3~keV and Fe~K${\alpha}$ (E$\sim$6.4~keV) and Fe~K${\beta}$ (E$\sim$7~keV) emission lines and Fe absorption edge (E$\sim$7.2~keV) significantly vary among the whole data set. 

Based on the light-curve analysis and hardness ratio evolution found, for observation \#74 we extracted three spectra, each corresponding to the eclipse (a), egress (b) and out-of-eclipse (c) periods described in \hyperref[sec:gti]{Section~\ref{sec:gti}}. For each of the rest of the observations we extracted a single time-averaged spectrum, since no significant hardness-ratio variations were found. The eight resulting spectra along with the best fit models and their residuals are shown in \hyperref[fig:specs]{Figure~\ref{fig:specs}}. 

In order to model the spectra we used a doubly absorbed thermally Comptonized continuum. One of the absorption components is used to take into account the extinction associated to the interstellar medium (ISM) while the second one, a partial-absorber model, is used to model intrinsic absorption and soft excess present during ingress towards eclipse observation \#77. 
In the language of {\sc XSPEC}, the model reads: {\sc tbabs*pcfabs*nthcomp}. The ISM column density ({\sc tbabs} model) was fixed to a value\footnote{LAB weighted average obtained using {\em nh} tool from the {\sc FTOOL}s package.} of $N_{\rm H}$=0.8$\times10^{22}$~cm$^{-2}$ while the covering fraction $f$ ({\sc pcfabs} model) was fixed to 1 in all observations except for obs~\#77 and \#74b which show a large excess at lower energies. 
{\sc nthcomp} is the thermally Comptonized continuum model of \cite{Zdziarski1996} and extended by \cite{Zycki1999}. This model is parameterized by an asymptotic powerlaw index $\Gamma$, an electron temperature $kT_{\rm e}$, a seed photon energy $kT_{\rm bb}$, an input type for black-body (inp\_type$=$0) or disk black-body (inp\_type$=$1) seed photons. In our case, we chose a scenario with black-body type seed photons. Both high and low energy rollovers were kept fixed to their default values of 100~keV and 0.1~keV respectively because they are outside the extracted PN spectrum energy range. Gaussian profiles were used to model the emission lines present on each spectrum.

We set the cosmic elements abundances to those of \cite{Anders1989} and used photoelectric absorption cross-sections of \cite{Verner1996}. In order to estimate continuum and line fluxes we used the {\sc cflux} convolution model within {\sc xspec} package. The total unabsorbed luminosity was calculated using a distance of 12.4~kpc. Errors in all parameters were calculated within the 90\% uncertainty level using Markov Chain Monte Carlo ({\sc chain} task in {\sc xspec}). We set it up with the Goodman-Weare algorithm using 8 walkers and $10^5$ steps each.

In \hyperref[fig:specs]{Figure~\ref{fig:specs}} we show the eight spectral fits with the lower panel indicating $\chi^2$ deviations. 
Continuum parameters and fluxes are shown on \hyperref[tab:cont]{Table~\ref{tab:cont}} and the emission lines parameters and fluxes on \hyperref[tab:lines]{Table~\ref{tab:lines}}.


\begin{table*}
\centering
\begin{tabular}{c c | c c c c c c c}
\hline
\hline
OBSID  & Orbital Phase & $N_{\rm H}$            & $\Gamma$               & Flux                     & Luminosity           & $\chi^{2}$/dof \\ 
   & & 10$^{22}$~cm$^{-2}$    & --                     & 10$^{-12}$~erg/s/cm$^2$  &    10$^{36}$~erg/s   &  --              \\
\hline
76  & -0.376$\pm$0.019 & 5.48$^{+0.31}_{-0.23}$  & 0.55$^{+0.05}_{-0.04}$  & 95.5$^{+1.5}_{-1.5}$  & 1.76$^{+0.03}_{-0.03}$    & 1.39 / 174  \\
77  & -0.155$\pm$0.016 & 36.06$^{+3.66}_{-2.02}$ & 0.27$^{+0.10}_{-0.11}$  & 55.0$^{+4.0}_{-3.9}$  & 1.01$^{+0.07}_{-0.07}$    & 1.03 / 97   \\
74a &  0.074$\pm$0.018 & 0.17$^{+0.97}_{-0.14}$  & 0.02$^{+0.27}_{-0.14}$  & 1.0$^{+0.1}_{-0.1}$   & 0.019$^{+0.002}_{-0.002}$ & 1.19 / 23   \\
74b &  0.109$\pm$0.018 & 14.70$^{+0.99}_{-0.8}$  & 0.66$^{+0.07}_{-0.06}$  & 54.5$^{+2.8}_{-2.5}$  & 1.00$^{+0.05}_{-0.05}$    & 1.13 / 61 \\
74c &  0.144$\pm$0.018 & 5.99$^{+0.35}_{-0.29}$  & 0.86$^{+0.09}_{-0.06}$  & 61.1$^{+1.6}_{-1.5}$  & 1.13$^{+0.03}_{-0.03}$    & 1.10 / 135  \\
26  &  0.188$\pm$0.009 & 8.0$^{+0.7}_{-0.5}$     & 0.76$^{+0.10}_{-0.07}$  & 65.7$^{+2.00}_{-1.8}$ & 1.21$^{+0.04}_{-0.03}$    & 1.09 / 82   \\
78  &  0.237$\pm$0.008 & 0.99$^{+0.08}_{-0.07}$  & 1.01$^{+0.04}_{-0.04}$  & 75.7$^{+1.1}_{-1.1}$  & 1.39$^{+0.02}_{-0.02}$    & 0.90 / 155  \\
75  &  0.377$\pm$0.018 & 2.25$^{+0.09}_{-0.13}$  & 0.79$^{+0.03}_{-0.04}$  & 57.4$^{+0.7}_{-0.7}$  & 1.06$^{+0.01}_{-0.01}$    & 1.09 / 168  \\

\hline
\end{tabular} 
\caption{Best-fit parameters for the continuum model of the extracted spectra. Both unabsorbed {\sc nthcomp} flux and luminosity were calculated between 0.5--12~keV. We adopted a luminosity distance of 12.4~kpc \citep{Torrejon2010}. }
\label{tab:cont}
\end{table*}


Fe~K$\alpha$ ($\sim$6.4~keV) and Fe~K$\beta$ ($\sim$7~keV) emission lines were fitted on every observation, although in some of them they were not clearly resolvable. We tested for the presence of Fe~K$\alpha$ and Fe~K$\beta$ emission lines by running simulations using the {\em fake-it} task from {\sc xspec}. We fitted each line normalization (with fixed line energy and width) and then drawn $10^5$ sampled spectra with the same model and continuum parameters and proceeded to count how many of the sampled spectra had a line normalization greater than the normalization fitted to the real spectra. The ratio $r$ between the latter number and the total sampled spectra gives an estimate of the probability of getting a normalization higher than the real one by chance. The smaller this ratio $r$, the greater the confidence with which we can assume the actual presence of an emission line. We present this ratios as $1-r$ in \hyperref[tab:lines]{Table~\ref{tab:lines}} in percentage units. 

 Line energies, widths and fluxes were computed according to this table by establishing a lower tolerance of $\sim$99\% (1--r $>$ 0.99). On some cases line widths were left frozen and/or only computed an upper limit to fluxes. These are shown with an $\dagger$ symbol on \hyperref[tab:lines]{Table~\ref{tab:lines}}.

\begin{table*}
\centering
\begin{tabular}{c c | c c c c | c c c c }
\hline
\hline
OBSID & Orbital Phase & DP & $E$ & $\sigma$ & Flux  & DP & $E$ & $\sigma$ & Flux \\
        & & \%  & keV    &  keV   & 10$^{-12}$~erg/s/cm$^2$   &  \%  & keV    &   keV    & 10$^{-12}$~erg/s/cm$^2$ \\
\hline
76   & -0.376$\pm$0.019 & 100 & 6.37$^{+0.02}_{-0.02}$ & 0.06$^{+0.04}_{-0.05}$ & 1.2$^{+0.2}_{-0.2}$ & 94.2 & -- & -- & -- \\
77   & -0.155$\pm$0.016 & 100 & 6.42$^{+0.01}_{-0.02}$ & 0.06$^{+0.02}_{-0.05}$ & 1.3$^{+0.4}_{-0.5}$ & 100 & 6.91$^{+0.05}_{-0.06}$ & 0.21$^{+0.15}_{-0.003}$ & 1.4$^{+0.8}_{-0.4}$ \\
74a  & 0.074$\pm$0.018 & 100 & 6.49$^{+0.07}_{-0.04}$ & 0.16$^{+0.05}_{-0.04}$ & 0.16$^{+0.03}_{-0.03}$ & 100 & 6.89$^{+0.15}_{-0.26}$ & 0.12$^{+0.21}_{-0.1}$ & 0.06$^{+0.09}_{-0.04}$ \\
74b  & 0.109$\pm$0.018 & 100 & 6.41$^{+0.09}_{-0.08}$ & 0.1$^{\dagger}$ & 0.38$^{+0.16}_{-0.16}$ & 97.7 & 6.92$^{+0.09}_{-0.1}$ & 0.1$^{\dagger}$ & 0.41$^{+0.17}_{-0.17}$ \\
74c  & 0.144$\pm$0.018 & 45.5 & 6.4$^\dagger$ & 0$^\dagger$ & $<$0.13 & 89.4 & -- & -- & -- \\
26   & 0.188$\pm$0.009 & 98.6  & 6.4$^\dagger$ & 0.1$^\dagger$ & 0.23$^{+0.15}_{-0.14}$ & 97.9 & -- & -- & -- \\
78   & 0.237$\pm$0.008 & 100 & 6.39$^{+0.08}_{-0.07}$ & 0.10$^{+0.18}_{-0.07}$ & 0.6$^{+0.2}_{-0.2}$ & 52.4 & -- & -- & -- \\
75   & 0.377$\pm$0.018 & 100 & 6.40$^{+0.02}_{-0.03}$ & 0.02$^{+0.07}_{-0.01}$ & 0.63$^{+0.12}_{-0.09}$ & 99.1 & 6.94$^{+0.22}_{-0.03}$ & 0.045$^{+0.16}_{-0.04}$ & 0.22$^{+0.09}_{-0.1}$ \\

\hline
\end{tabular} 
\caption{Fe~K$\alpha$ and Fe~K$\beta$ detection probabilities (DP), line energies ($E$), widths ($\sigma$) and unabsorbed fluxes. $\dagger$ symbol indicates the corresponding parameter was frozen.}
\label{tab:lines}
\end{table*}


As can be seen from the obtained $\chi^2$ statistics, good overall fits were obtained from the {\sc nthcomp} model. Pre-eclipse observation \#77 large soft excess at energies lower than 3~keV, and thus was fitted with a free covering fraction. The covering fraction for this case gives 0.95$\pm$0.01.
The continuum shape remains very similar throughout the rest of the non-eclipsing observations (\#75, \#76 and \#78), indicating that the NS accretion rate and wind density does not vary much along the orbit.

The spectral index becomes harder ($\ll$1) on observations \#77 and \#74b, which also show an increase in the X-ray absorption, a decrease on X-ray fluxes and hardness-ratio rising. Among all the out-of-eclipse observations, X-ray emission becomes softer, with indeces closer to $\sim$1, which are typical values for this kind of systems.

Outside the eclipse, \igr unabsorbed luminosities oscillate between $\sim$1.0~L$_{36}$, just exiting the eclipse, and $\sim$1.8~L$_{36}$, according to its orbital phase (L$_{36}$=10$^{36}$~erg/s). During the eclipse egress (\#74b), the source luminosity increases by a factor of $\sim$50 with respect to the eclipse (\#74a).

From the emission line perspective, observations \#26 and \#74c look featureless. \citet{Hill2005} made an F--test on observation~\#26 to check the presence of Fe~K$\alpha$ and could not confirm the line presence within 90\% confidence region. According to \hyperref[tab:lines]{Table~\ref{tab:lines}}, we can detect the Fe~K$\alpha$ line on observation \#26 with a significance of $\lesssim$99\% while on observation \#74c with a $\lesssim$46\% significance. Thus, in both these cases we could give an upper limit to the line flux with its energy and width fixed.

Although without emission lines, these two observations present an absorption edge around $\sim$7.2~keV. To better describe it we added an absorption {\sc edge} convolution model to their respective continuum. The results were consistent within the two observations, giving a threshold energy of 7.15$^{+0.07}_{-0.08}$~keV and absorption depth of 0.23$\pm$0.09. 

\begin{figure*}
\centering
\includegraphics[width=0.49\textwidth,clip]{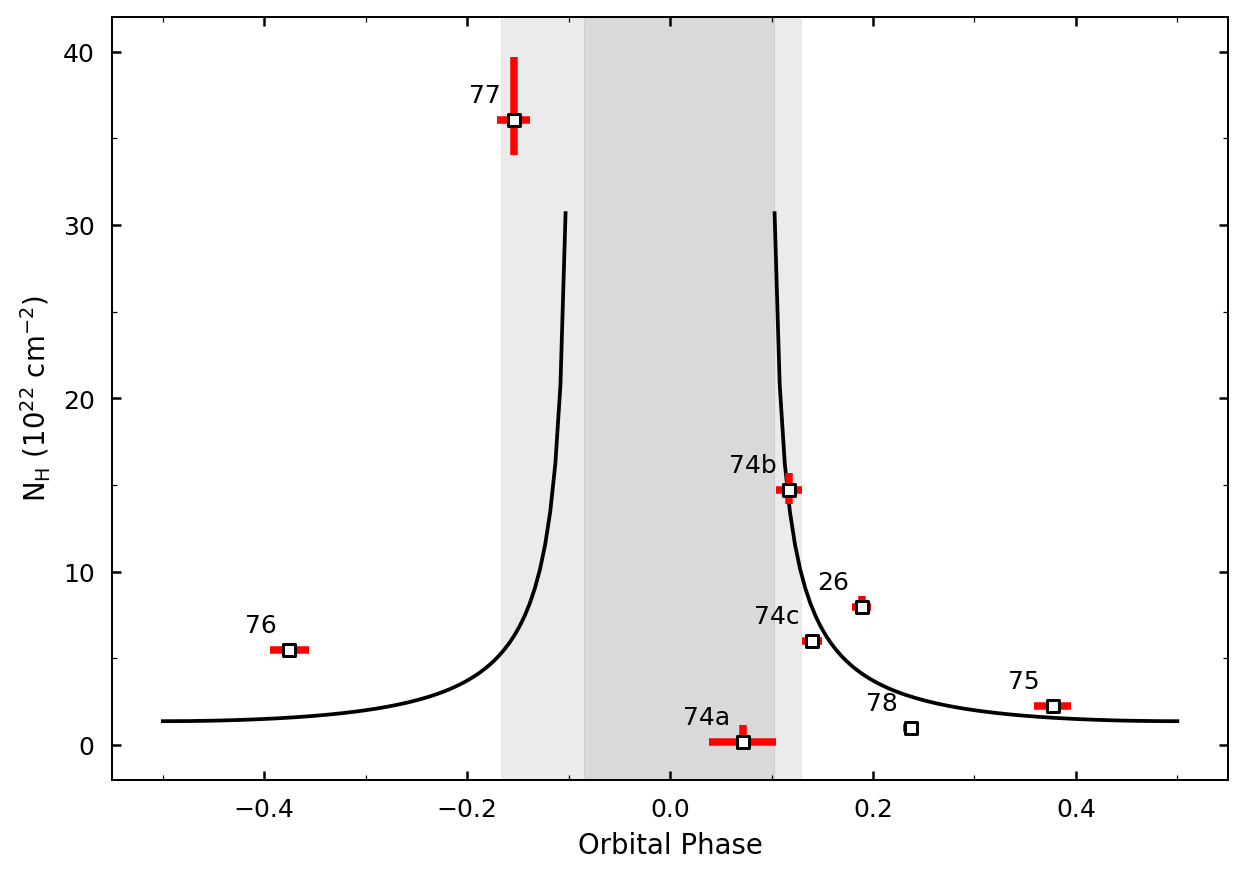}
\includegraphics[width=0.49\textwidth,clip]{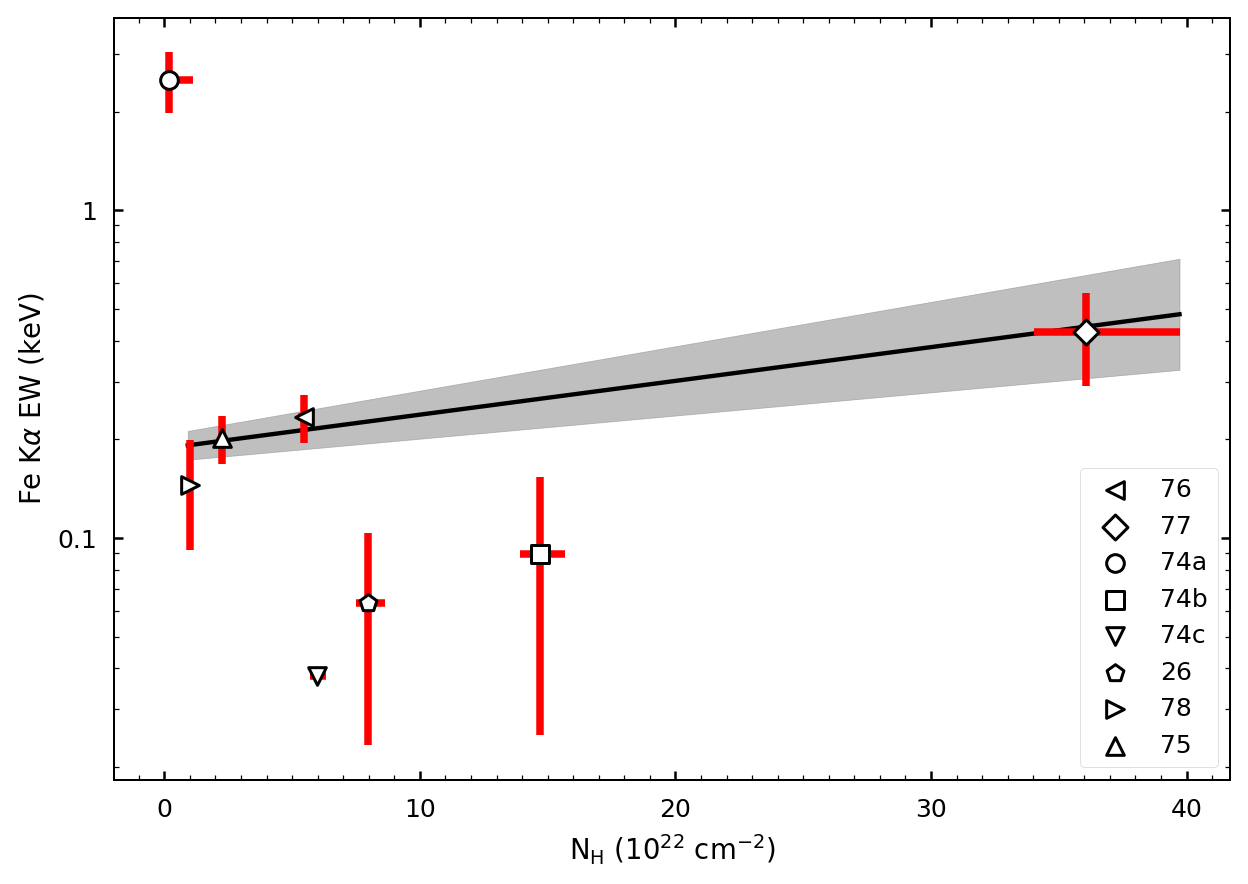}
\caption{
Left: Phase evolution of the absorption column density. Light grey colour stripes indicate ingress and egress transitions, while dark grey indicates the eclipse. Full black line corresponds to a \citet[][]{Castor1975} wind model fit using system parameters from \citet{Coley2015}. A clear excess over the model can be seen for phases previous to the eclipse suggesting an additional contributing component than just wind absorption.
Right: Fe~K$\alpha$ unabsorbed equivalent width against the absorption column. Labels are phase ordered. The linear behaviour in non-eclipsing phases suggests that the Fe~line emission region follows closely the NS orbital motion.
}
\label{fig:nheqwidth}
\end{figure*}

Observations \#77, \#74a and \#74b present strong Fe emission lines and absorption edges. This effect comes from the fact that continuum emission is very diminished with respect to the other observations, giving also a higher equivalent width. Observations \#77 and \#74b present also the highest measured values of absorption, the largest one being prior to the eclipse (see left panel of \hyperref[fig:nheqwidth]{Figure~\ref{fig:nheqwidth}}).

In order to study the Fe~K$\alpha$ flux behaviour along the orbit we computed its equivalent width (EW) for each observation and compared it with the absorption column. The EW was calculated by computing the ratio between the unabsorbed Fe~K$\alpha$ flux and the unabsorbed continuum flux, both between 6.1--6.7~keV. This plot is shown on the right panel of \hyperref[fig:nheqwidth]{Figure~\ref{fig:nheqwidth}}.

Outside the eclipse the EW seems to follow a positive linear correlation. We made a fit excluding low-significance data (Obs.~\#74 and \#26) which resulted in a slope of 0.01$\pm$0.003 with a reduced $\chi^2\sim$~0.54. This result is in agreement with that of \cite{Inoue1985}, which gave an approximate relationship between these two quantities: 
\begin{equation}
	{\rm EW} \approx 100~{\rm eV} \times N_{23}
\end{equation}
where $N_{23}$ is the absorption column in units of $10^{23}$~cm$^{-2}$. This expression corresponds to the case when the NS is totally hidden by a dense cocoon of matter and only scattered emission from the ambient matter is observed.





\begin{figure}
\centering
\includegraphics[width=\columnwidth,clip]{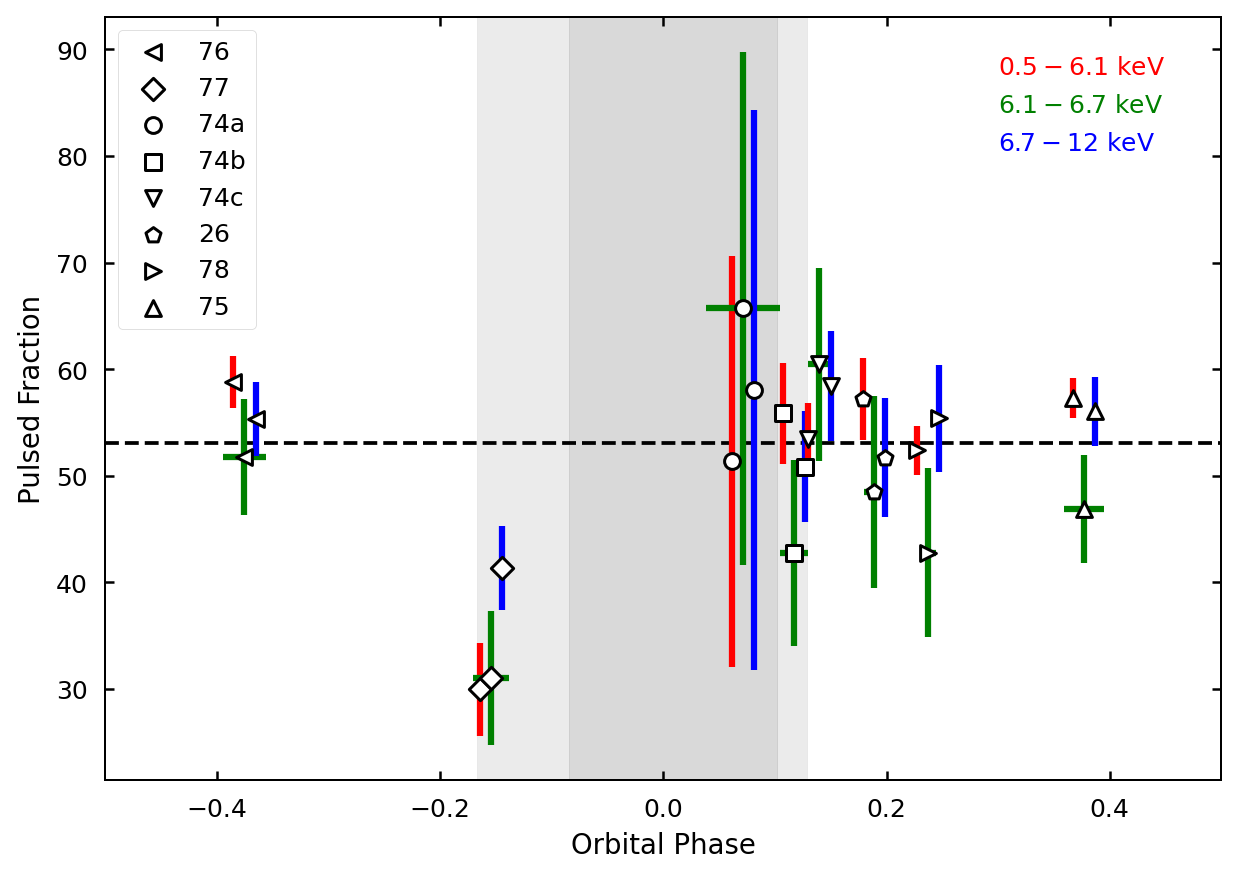}
\caption{Pulsed fraction (definition in text) for each observation calculated in 3 energy bands. Light curves used were folded with the best spin period found according to \hyperref[fig:spins]{Figure~\ref{fig:spins}}. Dashed horizontal line corresponds to the out of eclipse average value of 53$\pm$5 \%.}
\label{fig:pfraq}
\end{figure}

We also analyzed the orbital evolution of the NS pulsed fraction ($PF$), shown on \hyperref[fig:pfraq]{Figure~\ref{fig:pfraq}}. The NS pulsed fraction was calculated by adopting the following definition:
\begin{equation}
	PF = \frac{F_{\rm max}-F_{\rm min}}{F_{\rm max}+F_{\rm min}}
\end{equation}
where F$_{\rm max}$ and F$_{\rm min}$ are the maximum and minimum values of the folded lightcurve (1~s bin on the given energy band) using the best-period found for the respective observation according to \hyperref[fig:spins]{Figure~\ref{fig:spins}}. We calculated the $PF$ for each observation on 3 different energy bands: 0.5--6.1~keV, 6.1--6.7~keV and 6.7--12~keV.

We note that for every other non-eclipse observation the $PF$ varies around an average value of 53$\pm$5\% (dashed horizontal line on \hyperref[fig:pfraq]{Figure~\ref{fig:pfraq}}) with a soft X-ray ($<$6.1~keV) dominating component. But for the entering-eclipse observation \#77, the total $PF$ remains statistically below the averaged value and dominated by harder X-rays ($>$6.7~keV). 


\section{Discussion} \label{sec:discussion}

The classical picture of sgXBs involve a compact object (CO), typically a neutron star (NS) in a close orbit around a supergiant early star which has a strong wind embedding it and from which the CO is constantly accreting. By considering a simple spherical wind \citep[Castor-Abbott-Klein (CAK),][]{Castor1975}, one can model the absorption orbital modulation by integrating the wind density profile along the line of sight for each orbital phase \citep[see][for a full description on this procedure]{Garcia2018}. However, for a very low eccentricity binary like \igr, this method would produce a symmetrical absorption profile with respect to the eclipse mid-phase, which is in disagreement with the asymmetric absorption eclipse profile from \igr, as shown in \hyperref[fig:nheqwidth]{Figure~\ref{fig:nheqwidth}}. Since this asymmetry persists through several orbital revolutions, it is natural to invoke the presence of an asymmetric component that could explain the increase of material between the observer and the NS when the latter is entering the eclipse, and which has to be behind it when it appears back from the supergiant star at the eclipse egress. 

\begin{figure}
\centering
\includegraphics[width=\columnwidth, clip]{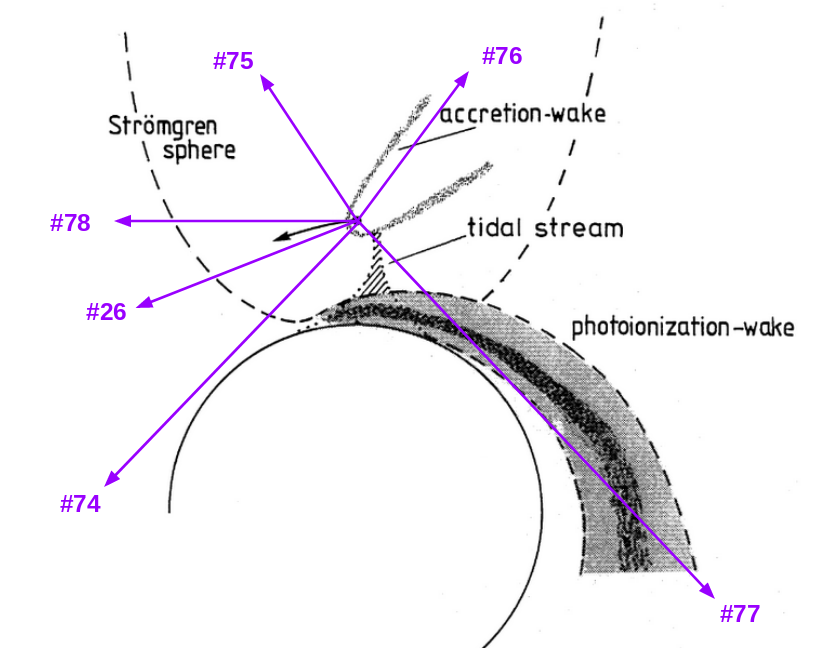}
\caption{ Schematic representation of the model scenario. The \pwake is responsible for the increased absorption column asymmetry before the eclipse. An \awake might also be present and increase absorption in phases previous to that of the eclipse.  Figure adapted from \citet{Kaper1994}.}
\label{fig:model}
\end{figure}

The high absorption measured at the proximity of the eclipse (when the line of sight between the NS and the observer is almost tangent to the companion star), and the fact that outside the eclipse the absorption column is a factor of 3--10 times lower (remaining stable within errors), cannot be explained only by a symmetric wind emanating from the supergiant companion. An over density of absorbing matter, intersecting the line of sight towards the NS and enhancing the absorption column before the eclipse, is needed to explain both the asymmetry during the eclipse and the absorption columns measured in observations \#77 and \#74b. This asymmetry can be explained by considering a scenario where a \pwake is present (see \hyperref[fig:model]{Figure~\ref{fig:model}}). This density feature arises from the collision of the undisturbed radiation driven wind and the stagnant highly ionised plasma inside the NS \stromgren \,zone, producing strong shocks that create a dense sheet of gas that trails the NS. This kind of shock has also been proposed to explain asymmetric absorption profiles and hard X-ray light curves behaviour in other eclipsing HMXBs such as Vela~X--1 and 4U~1700--37 \citep{Kaper1994, Feldmeier1996}. 

Observational evidence for the presence of dense slabs of material in the accretion flow of \igr are seen in the increase of X-ray hardness ratio towards eclipsing observations. Observations \#77 (ingress) and \#74b (egress) present an average hardness ratio 2--3 times higher than the observations far out-of-eclipse (see \hyperref[tab:colors]{Table~\ref{tab:colors}}). 
The emitted soft X-rays are partly absorbed by regions of higher densities in the line of sight towards the compact object at those orbital phases. The fact that the hardness ratio at the ingress to the eclipse is larger than that of the egress, suggest that the \pwake can account for this effect.

A couple of additional features are also expected due to the presence of the NS within the stellar wind : i. an \awake trailing the NS, resulting from the accretion of perturbed wind by the passage of the highly supersonic NS; and ii. a {\em tidal stream}, i.e. a stream of gas leaving the OB star and reaching the NS passing through the Lagrangian L1, formed by tidal interaction within the binary system. Both features are expected to be much less dense than the \pwake, and thus contribute weakly to the obscuration of the CO \citep{Blondin1990, Kaper2006}. This \awake can account for the absorption, higher than the wind only, seen before the ingress to the eclipse (Obs.~\#76; see \hyperref[fig:nheqwidth]{Figure~\ref{fig:nheqwidth}}). The slightly increased hardness seen in this observation (see \hyperref[tab:colors]{Table~\ref{tab:colors}}) can also contribute to this fact. More observations inbetween the phases $-0.5$ and $-0.2$ could help to determine the presence of an \awake more precisely. 
On the contrary, the {\em tidal stream} always being located inbetween the OB star and the NS, would not lead to an increased absorption. The lower pulse fraction seen in \hyperref[fig:pfraq]{Figure~\ref{fig:pfraq}} indicates a deviation from spherical symmetry in the ingress to the eclipse. Furthermore, as the $PF$ is statistically lower for energies $<$6.7~keV, this suggests that this asymmetry affects lower energy photons preferably, and hence supporting the presence of the \pwake.


These findings come in disagreement with what \cite{Pradhan2019} discuss in their recent paper on \igr using the same XMM-{\em Newton} data set (however, see our note about background and pile-up analysis in \hyperref[sec:reduction]{Section~\ref{sec:reduction}}). The authors model the absorption by means of two variable models: one for Galactic absorption and another for the local absorption. With this method, they obtain much lower local absorption column densities than we do (with increased Galactic absorption instead) and, although they obtain an increase of absorption before and after the eclipse, they argue that an \awake could not be formed on \igr based on the premise that the stellar wind velocity of this system is large enough to not allow the formation of such structures. 

The radius of influence of the X-ray pulsar is governed by its accretion radius and by the \stromgren \,sphere. It is expected that X-ray continuum emission should originate very close to the NS, generated by accretion of matter from the wind within these regions \citep{Kaper1994}. By means of the ionization parameter $\xi$ we can estimate the size of this X--ray emitting region. This parameter is defined as $\xi = L_X / (n * R^2)$ where $L_X$ is the illuminating X--ray luminosity, $n$ is the surrounding gas density and $R$ its distance to the X-ray source. Furthermore, the matter density can be related to the absorption column density $N_{\rm H}$ by $N_{\rm H}=n*R$. According to \cite{Inoue1985}, Fe ions up to Fe~XIX can contribute to the bulge of the 6.4~keV Fe line profile. Such ionization degree can be reached by ranges of $\xi$ between 10 and 300.

We can estimate the size $R$ of the emitting region, by taking into account the egress time interval ($\sim$10~ks) when the X--ray luminosity increases by a factor of 50 (see \hyperref[tab:cont]{Table~\ref{tab:cont}}), and using the NS orbital velocity. For the latter we used the same orbital parameters as for the $N_{\rm H}$ model shown in \hyperref[fig:nheqwidth]{Figure~\ref{fig:nheqwidth}}. Thus, we obtain a region radius of $\sim$1.9$\times$10$^{12}$~cm or equivalently, $\sim$27~R$_{\odot}$. If we now take into account the obtained luminosities and absorption columns from the spectral fits, we can give a range for the emitting-region size and density. Averaging $L_X$ and $N_{\rm H}$ for non-near eclipse observations, we get a region size between $10^{11}$~cm and $3\times10^{12}$~cm, or equivalently, between $\sim$1.4 and $\sim$43~R$_{\odot}$. Furthermore, this region has a matter density ranging between $1.5 \times 10^{10}$~cm$^{-3}$ and $4.4 \times 10^{11}$~cm$^{-3}$.

The fact that our first $R$ estimation lies very close to the higher end of the second estimation, suggests more firmly that Fe ions up to XIX might be present in the NS surroundings and contributing to the Fe~K$\alpha$ line complex. The positive linear correlation found between the unabsorbed Fe~K$\alpha$ equivalent width and the absorption column density (\hyperref[fig:nheqwidth]{Figure~\ref{fig:nheqwidth}}) and the low pulsed fraction seen in \hyperref[fig:pfraq]{Figure~\ref{fig:pfraq}} suggest that the Fe~K$\alpha$ line is generated close to the NS, and that this accreted matter is also responsible for the absorption. 

As we noted before, the Fe~K$\alpha$ line is generated in the vicinity of the NS. But the presence of the line during the eclipse indicates that these photons must come from another source, likely from reprocessing of X-ray photons by stellar wind particles, as already stated by \cite{Aftab2019}. This effect has also been seen in eclipsing binaries such as Vela X--1 where the line equivalent width during the eclipse is greater than 1 (as seen in \hyperref[fig:nheqwidth]{Figure~\ref{fig:nheqwidth}}.)



\section{Conclusions} \label{sec:conclusions}

In this work we have analyzed six archival XMM-{\em Newton} observations of the eclipsing HMXB \igr in addition to the 14 years cumulative hard X-ray {\em Swift}/BAT light curve. {\em Swift}/BAT folded light curve shows that this source has an asymmetric eclipse profile which spans a fraction of $\sim$0.2 of the total orbital cycle ($P_{\rm orb}\sim4.57$~days).

XMM-{\em Newton} light curves in the {\em soft} (0.5--6~keV) and {\em hard} (6--12~keV) energy bands show similar flaring behaviour compatible with NS stellar wind accretion as the origin of the X-ray emission. Observations \#74 and \#77 show a very low {\em soft} rate (higher {\em hardness} ratio) compared to other observations, as a consequence of the high obscuration occurring during eclipse ingress/egress. 

XMM-{\em Newton} time-averaged spectra show a highly absorbed power-law like continuum with Fe line and absorption features (Fe~K$\alpha$, Fe~K$\beta$ and Fe absorption edge) strongly dependent on the orbital phase. Observations outside the eclipse show an absorption column density lower than 10$\times$10$^{22}$~cm$^{-2}$ and Fe~K$\alpha$ emission line only. Eclipsing observations show a higher absorption column ($N_{\rm H}>30\times10^{22}$~cm$^{-2}$) and very strong Fe emission and absorption features. In particular, the absorption column before eclipse is $\sim$1.5 higher than that of the eclipse egress transition.

 The pulse fraction remains close to 50\% for all the observations except for that of the ingress \#77. For the latter it must be that the emitting matter does not co-rotate as much as the other observations with the NS spin. Moreover, photons with energies $<$6.7~keV are more heavily affected than those with greater energy, suggesting that this less-corotating matter is located on the line of sight, thus absorbing and scattering low energy photons.

 The Fe~K$\alpha$ equivalent width outside the eclipse follows a positive linear relationship with the absorption column. This indicates that the illuminated matter, where the Fe line originates, closely surrounds the NS, being also responsible for the absorption of the X-ray photons. The presence of the Fe line during the eclipse (EW$>$1) indicates that in this particular geometrical alignment, the emission line must be produced farther from the eclipsed NS, being likely formed in the stellar wind that surrounds the companion star.


Finally, in order to explain the asymmetric eclipse profile seen on the {\em Swift}/BAT hard X-ray light curve (\hyperref[fig:spins]{Figure~\ref{fig:spins}}), the absorption density column and hardness ratio orbital modulation (\hyperref[fig:nheqwidth]{Figure~\ref{fig:nheqwidth}} and \hyperref[tab:colors]{Table~\ref{tab:colors}}), we consider a \pwake trailing the NS, where matter is concentrated between the NS and the companion star. Accretion wake presence is also suggested by an increased absorption during pre-eclipse phases.


More observations of this source, specifically at phases before the eclipse and after inferior conjunction ($\phi < 0$) would be useful to better determine the presence and size of these large scale wind structures. Joint analysis with detailed numerical simulations (see e.g. \citet{Elmellah2020}) could provide great insight in the current picture of highly absorbed eclipsing sgHMXBs.

%
\begin{acknowledgements}
FAF, FG and JAC acknowledge support by PIP 0102 (CONICET). FAF is fellow of CONICET. JAC is CONICET researcher. This work received financial support from PICT-2017-2865 (ANPCyT). FG and SC were partly supported by the LabEx UnivEarthS, Interface project I10, ``From evolution of binaries to merging of compact objects''.
FG acknowledges the research programme Athena with project number 184.034.002, which is (partly) financed by the Dutch Research Council (NWO).
This work was partly supported by the Centre National d'Etudes Spatiales (CNES), and based on observations obtained with MINE: the Multi-wavelength INTEGRAL NEtwork. 
JAC was also supported by grant PID2019-105510GB-C32/AEI/10.13039/501100011033 from the Agencia Estatal de Investigaci\'on of the Spanish Ministerio de Ciencia, Innovaci\'on y Universidades, and by Consejer\'{\i}a de Econom\'{\i}a, Innovaci\'on, Ciencia y Empleo of Junta de Andaluc\'{\i}a as research group FQM-322, as well as FEDER funds.


\end{acknowledgements}

%

\begin{appendix}
\section{\xmm PN background treatment}
\label{sec:appendix}
In this section we describe the complete background treatment applied to the PN dataset. This full description is important on this context to avoid introducing systematic errors on the lightcurves and spectra.

\begin{figure*}
\centering
\includegraphics[width=\columnwidth,clip]{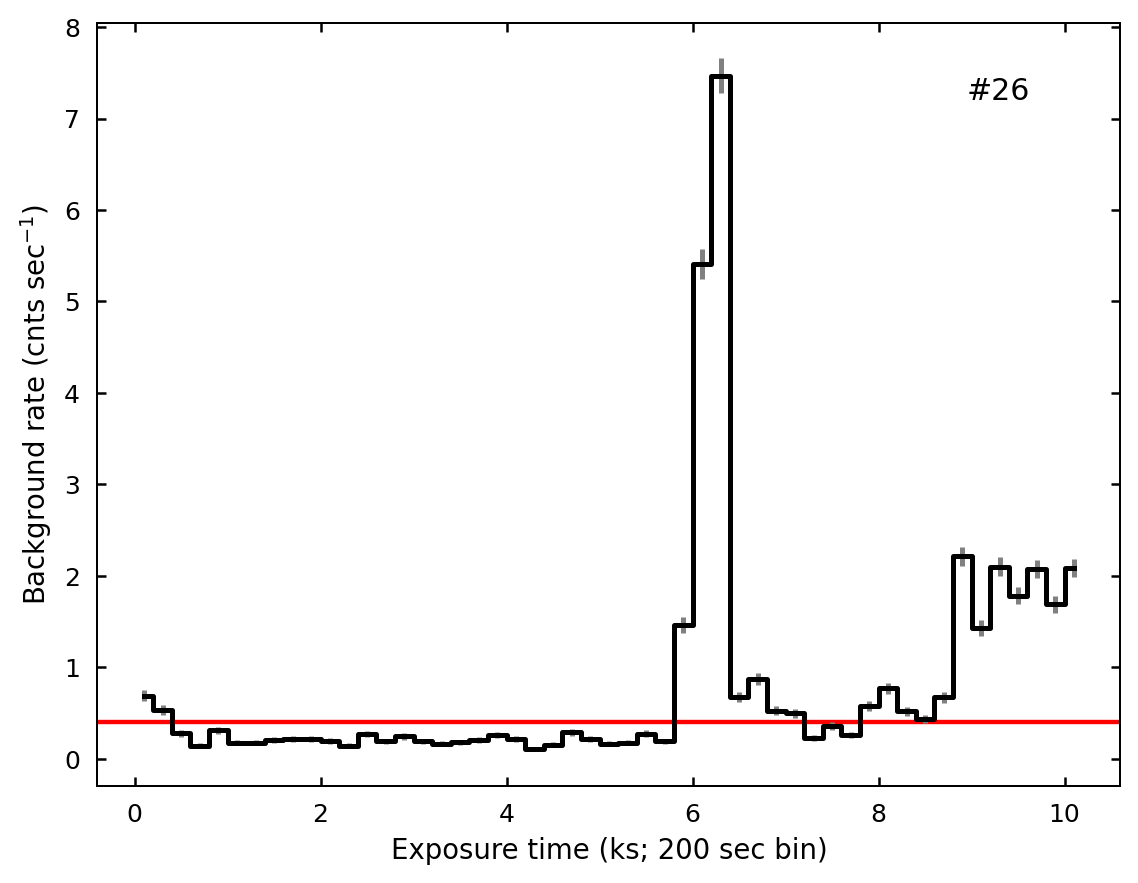}
\includegraphics[width=\columnwidth,clip]{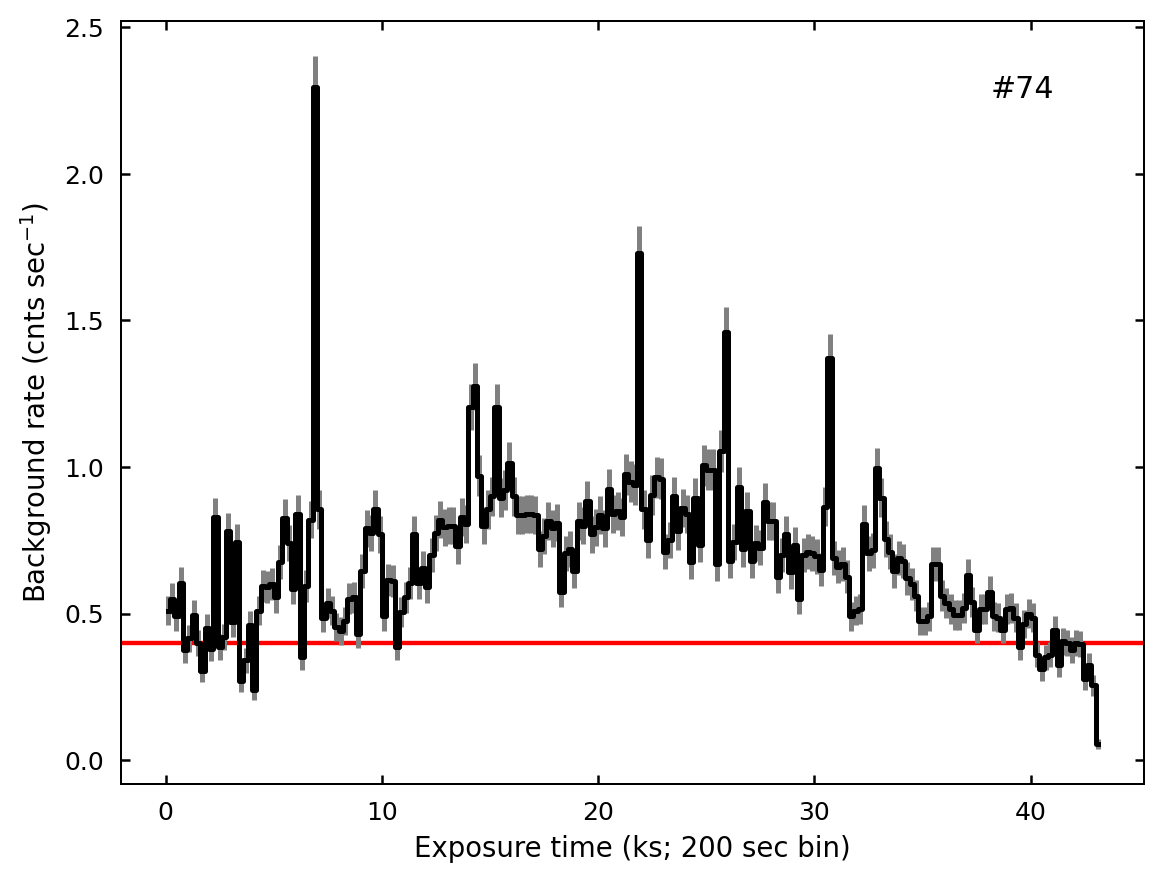}
\includegraphics[width=\columnwidth,clip]{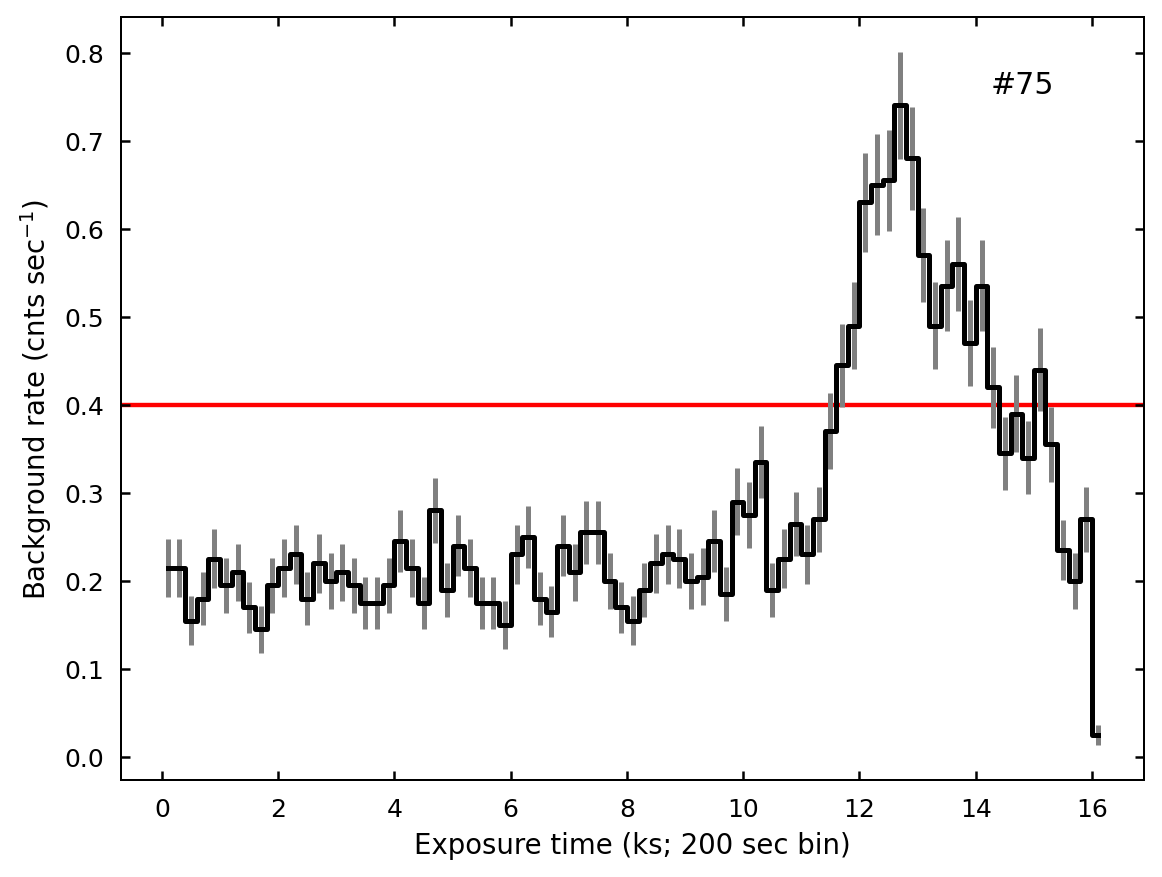}
\includegraphics[width=\columnwidth,clip]{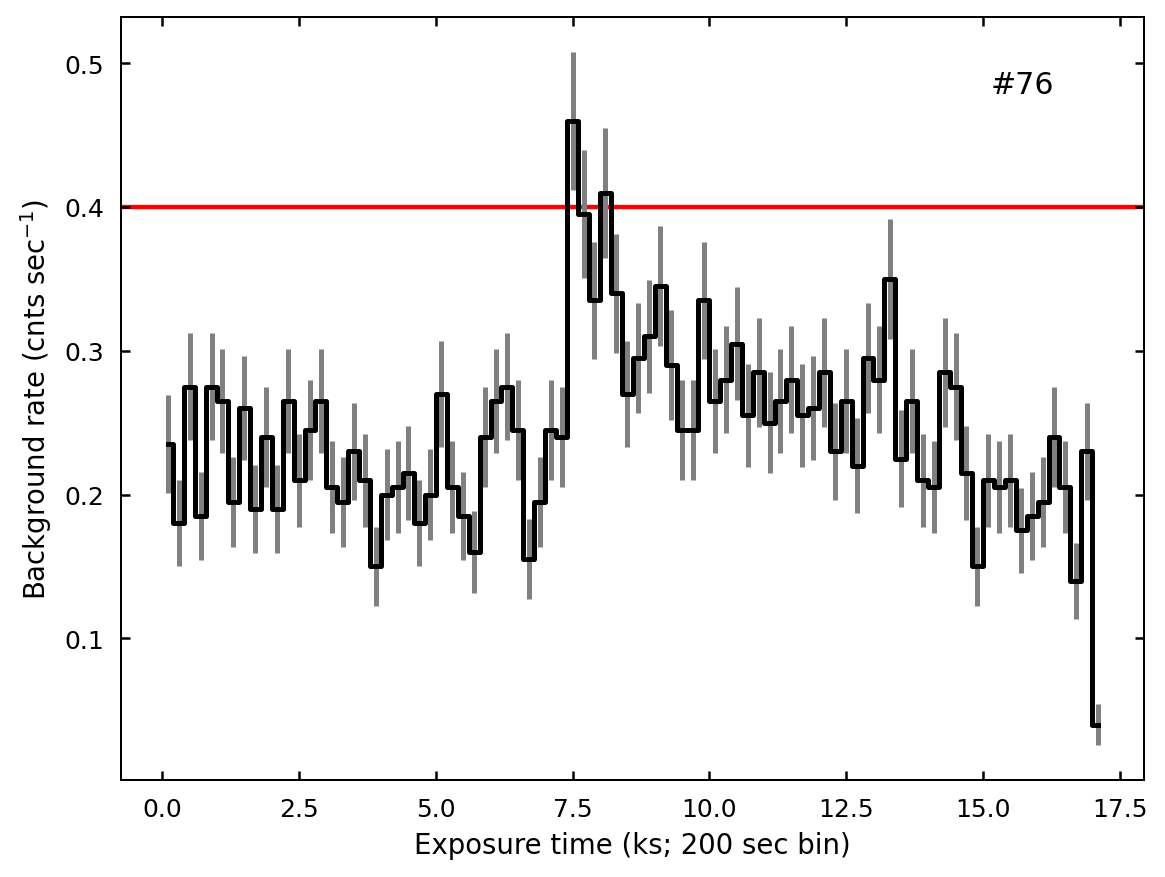}
\includegraphics[width=\columnwidth,clip]{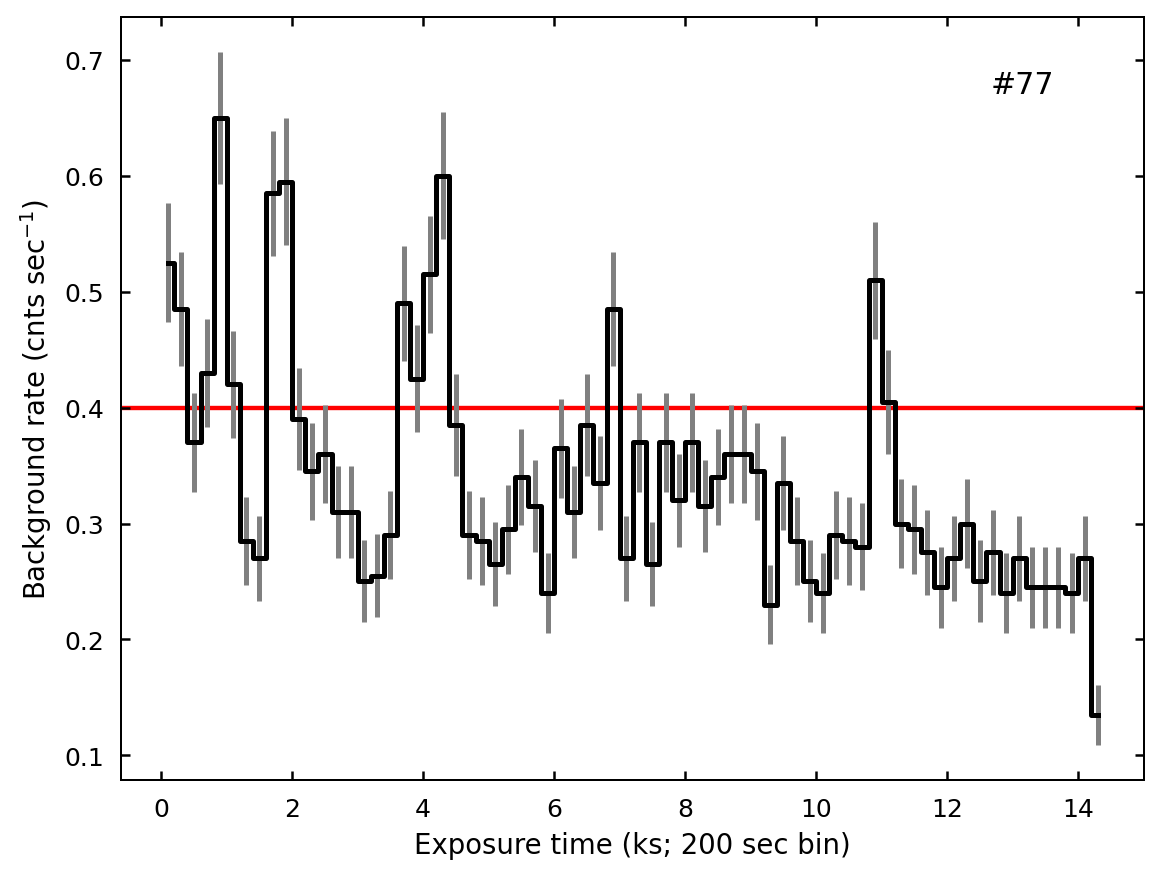}
\includegraphics[width=\columnwidth,clip]{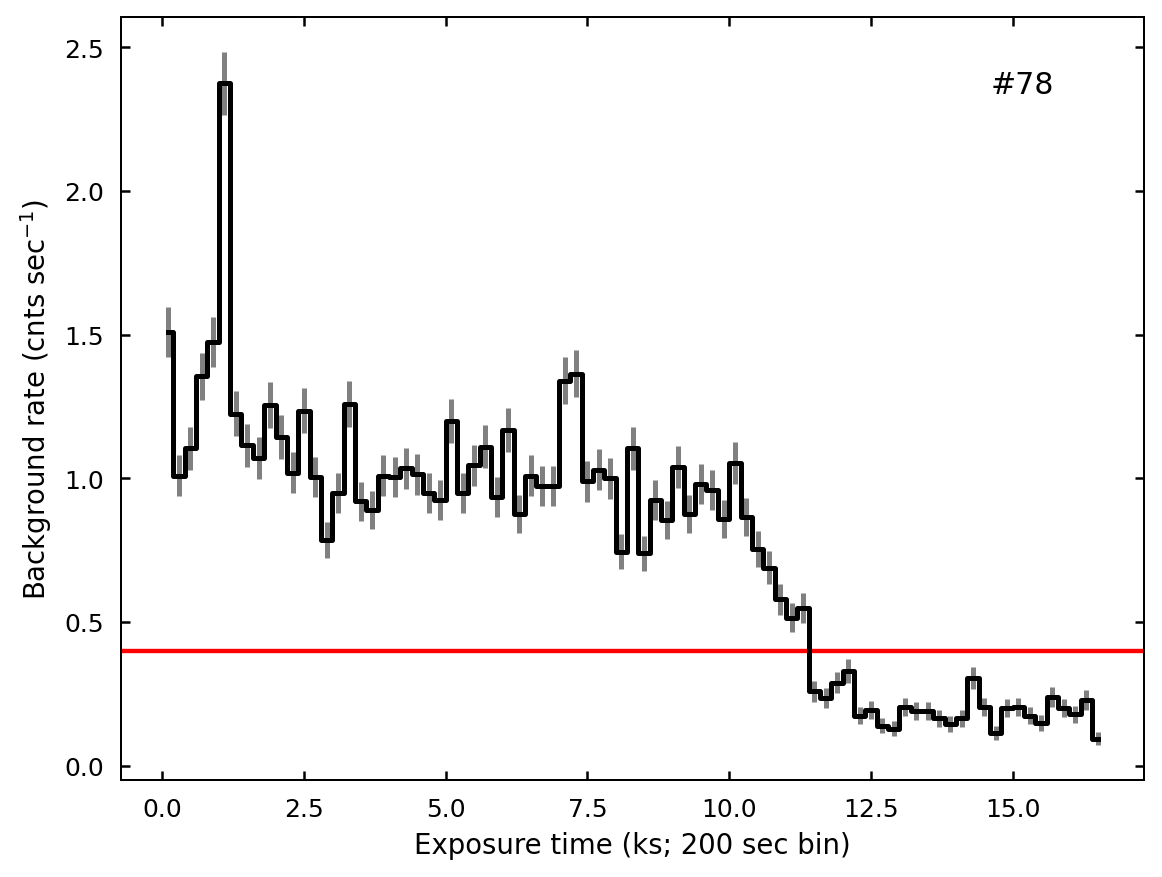}
\caption{\xmm PN high energy (E$>$10~keV) single events ({\sc PATTERN}==0) background (excluding an enlarged source region) light curves with a 200~second bin size. Standard 0.4~count rate limit is indicated with the horizontal red line. }
\label{fig:pnbkglcs}
\end{figure*}

On \hyperref[fig:pnbkglcs]{Figure~\ref{fig:pnbkglcs}} we show the entire collection of background lightcurves. These were extracted using only high-energy ($E>10$~keV) single events ({\sc PATTERN}==0). We also excluded a circular region of radius 1200~PhU, surrounding the source \igr . The standard count-rate limit of 0.4~cps of the PN camera is indicated in each panel with a red-horizontal line. As already described on \hyperref[sec:gti]{Section~\ref{sec:gti}}, eclipse observation \#74 presents a very high background activity but is not the only one. Obs~\#26, \#75 and \#78 also present flaring periods of several kiloseconds.
In contrast, \cite{Pradhan2019} only report Obs~\#26 presenting flaring activity.

\begin{figure*}
\centering
\includegraphics[width=\columnwidth,clip]{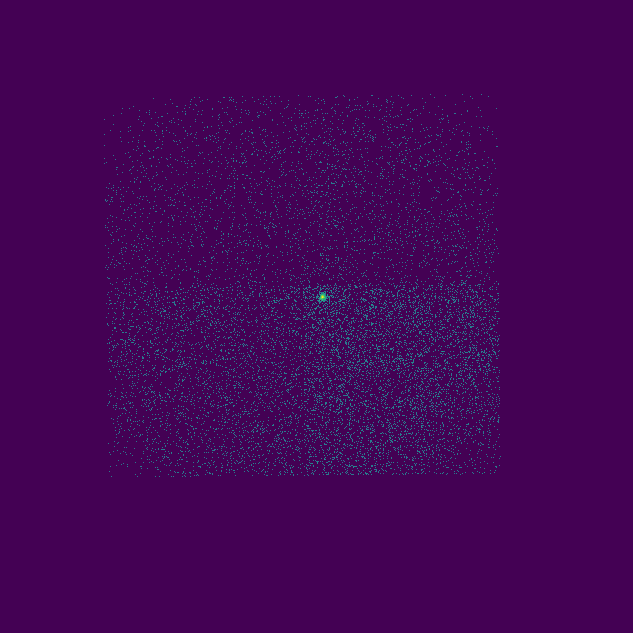}
\includegraphics[width=\columnwidth,clip]{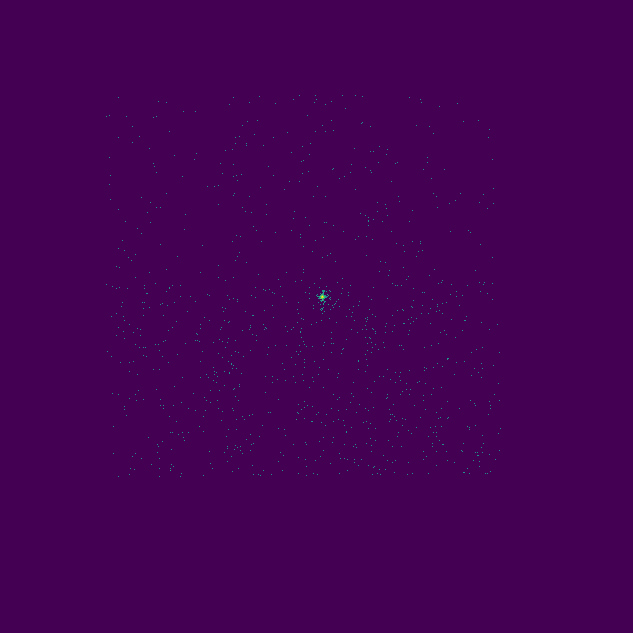}
\caption{\xmm PN high energy (E$>$10~keV) single events (PATTERN==0) images of observation \#78. Left panel corresponds to first 10ks of the background lightcurve ($<$rate$>\sim$1.1~cps). Right panel the corresponds to the remaining exposure starting at 12ks ($<$rate$>\sim$0.2~cps). }
\label{fig:pnbkgimage}
\end{figure*}

In order to show these differences more clearly, we extracted images of the whole CCD with the same background lightcurves configuration but for high and low activity periods seperately.
On \hyperref[fig:pnbkgimage]{Figure~\ref{fig:pnbkgimage}} we show an example for observation~\#78. Left panel corresponds to the first 10~ks of the observation where the average rate is $\sim$1.1~cps, while the right panel to the last 5~ks where the average rate decreaces to $\sim$0.2~cps. Comparing both panels it can be seen that this high activity background periods iluminate the entire CCD, so no background region selected is going to be free from contamination. Thus, if not removed accordingly, all further spectra and lightcurves will be systematically hardened.

\end{appendix}

\end{document}